\documentclass[aps,prl,reprint,preprintnumbers,nofootinbib,superscriptaddress]{revtex4-2}
\input{header.tex}

\begin{document}

\preprint{CERN-TH-2026-187, TU-1327}

\title{Heavy neutral leptons beyond the BBN bound: probing the lepton asymmetry of the Universe}

\author{Kensuke~Akita}
\email{kensuke.akita.e2@tohoku.ac.jp}
\affiliation{Department of Physics, Tohoku University, Sendai, Miyagi 980-8578, Japan}
\affiliation{Department of Physics, The University of Tokyo, Bunkyo-ku, Tokyo 113-0033, Japan}
\author{Maksym~Ovchynnikov}
\email{maksym.ovchynnikov@cern.ch}
\affiliation{Theoretical Physics Department, CERN, 1211 Geneva 23, Switzerland}
\author{Vsevolod~Syvolap}
\email{vsevolod.syvolap@csic.es}
\affiliation{Instituto de F\'isica Te\'orica UAM-CSIC, Calle Nicol\'as Cabrera 13-15, 28049 Madrid, Spain}

\date{\today}

\begin{abstract}
Hadronically decaying particles with lifetimes $\tau\gtrsim0.02\,{\rm s}$ are excluded by Big Bang Nucleosynthesis almost independently of their abundance: the mesons from their decays convert protons into neutrons faster than the reverse, and helium is overproduced. For heavy neutral leptons (HNLs), this blankets the couplings the upcoming accelerator searches can reach. We propose a scenario with large lepton flavor asymmetries in which HNLs evade it.  A Dirac HNL decays into $\pi^+$ and its antiparticle into $\pi^-$, so an asymmetry between the two populations injects an excess of $\pi^+$, which converts the neutrons back to their standard abundance, while a cancellation between the asymmetries and the HNL decays keeps the shift in $N_{\rm eff}$ small. This opens the parameter space with $m_l+m_\pi\lesssim m_N\lesssim1\,{\rm GeV}$ and $\tau_N\lesssim1\,{\rm s}$, including a substantial region within the reach of SHiP. A discovery there, combined with the BBN and CMB observables, would fix the primordial flavor asymmetries up to one remaining direction.  The relic neutrino background and gravitational waves from the QCD transition epoch may probe that direction, and with it the lepton asymmetry of the Universe.
\end{abstract}

\maketitle

\begin{figure}[t!]
    \centering
    \includegraphics[width=\linewidth]{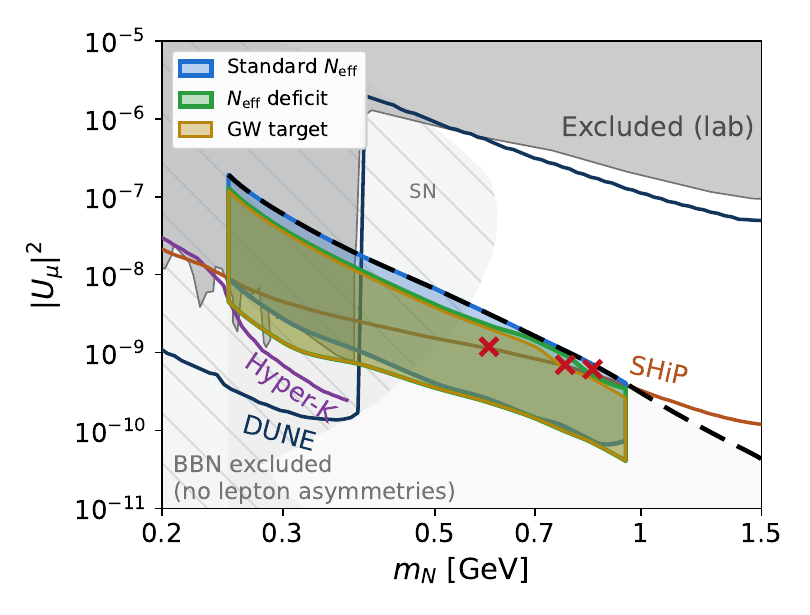}
    \caption{Parameter space of HNLs mixing with $\nu_\mu$: laboratory exclusions (gray); DUNE, SHiP, and Hyper-K sensitivities~\cite{deBlas:2025gyz}; the exclusion with no lepton asymmetry (pale, below the black dashed line); and the optimistic supernova constraint of Ref.~\cite{Carenza:2023old} (gray hatching).  Colored regions show the parameter space opened by the scenario with large lepton flavor asymmetries; the standard $\neff$ (blue), $\neff$ deficit (green), and GW target (gold) regions are defined in Table~\ref{tab:domain-definitions}; the GW target region marks where the QCD plasma enters the pion-condensed phase, a target for gravitational wave searches (SuM~\ref{app:qcd}), and lies inside the $\neff$ deficit region.  Red crosses mark the benchmarks of Table~\ref{tab:benchmarks}.}
    \label{fig:money}
\end{figure}

{\bf\textit{Introduction.}}--- Heavy neutral leptons (HNLs)~\cite{Abdullahi:2022jlv}, i.e., singlet fermions $N$ mixing with active neutrinos $\nu_\alpha$, are among the primary targets of the upcoming intensity frontier: SHiP, with complementary sensitivity from DUNE and Hyper-K at lower masses, is designed to probe mixings $U^2_\alpha$ orders of magnitude below current bounds for masses $m_N\sim0.1$ to $2\gev$~\cite{deBlas:2025gyz,SHiP:2025ows}. Standard cosmology, however, appears to forbid the domain of small couplings and long lifetimes (Fig.~\ref{fig:money}).  This covers the smallest mixings SHiP would reach below $m_N\simeq1\gev$, and almost the entire reach of DUNE and Hyper-K at $m_N\lesssim0.4\gev$.

In the early Universe, HNLs with $m_N>m_{l_\alpha}+m_\pi$ decay as $N\to\pi^+l^-_\alpha$, and the injected charged pions interconvert nucleons through strong reactions such as $\pi^{-}+p\to n+\pi^{0}/\gamma$, whose cross sections exceed the weak $\pn$ rates at MeV temperatures by sixteen orders of magnitude~\cite{Reno:1987qw,Kohri:2001jx,Pospelov:2010cw}. In the minimal scenario, decays of HNLs inject equal numbers of $\pi^+$ and $\pi^-$, and this drives the neutron fraction towards the isospin-symmetric value $n/p\simeq1$, overproducing helium and deuterium. Because the effect saturates, the resulting bound, $\tau_N\lesssim0.02\s$, depends on the abundance only logarithmically and is nearly universal for any hadronically decaying relic~\cite{Kawasaki:2017bqm,Fradette:2018hhl,Boyarsky:2020dzc,EscuderoAbenza:2025tsi,Jung:2025dyo}.  Without changing HNL interactions~\cite{Deppisch:2024izn,Dev:2025pru} (and, in particular, laboratory limits), the only known escape was to suppress the population itself through low-scale reheating~\cite{Gelmini:2004ah,Gelmini:2008fq}.

In this work, we identify and test, to our knowledge, the first mechanism by which a population of hadronically decaying particles produced by mixing can leave the BBN observables inside their observational windows, in the presence of large lepton flavor asymmetries. Several models generate such asymmetries~\cite{March-Russell:1999hpw,Kawasaki:2002hq,Yamaguchi:2002vw,Gu:2010dg,Mukaida:2021sgv,Kawasaki:2022hvx,Kasai:2024diy,Akita:2025zvq}, which affect BBN, the CMB, the production of sterile neutrino dark matter, and the QCD transition~\cite{Froustey:2024mgf,Domcke:2025lzg,Domcke:2025jiy,Gorbunov:2025nqs,Vogel:2025aut,Kasai:2025xaw,Akita:2025txo,Gao:2021nwz,Vovchenko:2020crk,Ferreira:2025zeu}. Here, we show this for a Dirac HNL.  For BBN, the mechanism requires the charge of the injected pions to track the decaying HNL; GeV-mass $N$ injects $\pi^+$ and $\bar N$ injects $\pi^-$, since each charged pion is emitted together with the charged lepton carrying the parent's lepton number (Supplemental Material (SuM)~\ref{app:mechanism}). An asymmetry $L_N\equiv(n_N-n_{\bar N})/s$ then makes the decays inject unequal numbers of $\pi^+$ and $\pi^-$, and can return the neutron abundance to its standard value. For the CMB, the relaxation comes from the cancellation between the two impacts on $\neff$: the asymmetries raise it~\cite{Froustey:2024mgf}, while the HNL decays deplete it~\cite{Akita:2024nam,Ovchynnikov:2024rfu}.\footnote{Cosmology with large lepton flavor asymmetries and \emph{Majorana} HNLs was inspected in Ref.~\cite{Gelmini:2020ekg}, which did not include the meson-driven $\pn$ conversion from HNL decays.  In this mass range, that conversion is the dominant effect of HNLs on BBN, and the relaxation of the bounds reported there does not survive its inclusion (SuM~\ref{app:prior-works-comparison}).}

Detecting an HNL inside the naively excluded domain would signal a non-standard MeV-temperature Universe~\cite{Gelmini:2019esj,Hasegawa:2019jsa,Barbieri:2025moq}, and the asymmetry scenario ties such a discovery to correlated cosmological imprints.  BBN constrains the surviving HNL asymmetry; the CMB constrains $\neff$, with a negative shift potentially preferred by the current data~\cite{AtacamaCosmologyTelescope:2025nti,SPT-3G:2025bzu,Escudero:2026mgw} and within reach of ongoing observations~\cite{SimonsObservatory:2018koc}; the QCD plasma may enter a pion-condensed phase in a part of the selected parameter space, which could source a stochastic gravitational wave background.  The resulting regions are shown in Fig.~\ref{fig:money}; the relic $\nu$ sea is modified as well~\cite{Betti:2019ouf}.

\begin{table}[t!]
 \caption{Regions of Fig.~\ref{fig:money}.  Each $\Delta$ is the shift from standard BBN with no HNLs and no asymmetries.  All require $-0.0038\leq\Delta\yp\leq0.0013$  and $|\Delta(\mathrm{D/H})|/(\mathrm{D/H})\leq0.031$; the GW target region lies inside the $\neff$ deficit region.}
 \label{tab:domain-definitions}
 \begin{ruledtabular}
 \begin{tabular}{ll}  region & further requirement \\  \hline  standard $\neff$ & $|\dneff|\leq0.05$ \\  $\neff$ deficit & $-0.28\leq\dneff\leq-0.05$ \\  GW target & pion-condensed phase \\ \end{tabular}
 \end{ruledtabular}
\end{table}

\begin{table*}[t!]
\caption{Benchmark solutions from the domains of Fig.~\ref{fig:money}; $q_N$ and $\xi_{20}$ are defined in Eq.~\eqref{eq:diagnostics}.  All points keep $\yp$ and D/H standard;  the first and third realize the standard $\neff$ region, the second the $\neff$ deficit region.  The last column lists the experiments and signatures each point probes.}
\label{tab:benchmarks}
\centering
\begin{ruledtabular}
\begin{tabular}{ccccc}
$(m_N,\tau_N)$ & $q_N$ & $\xi_{20}$ & $\dneff$ & probes \\
\hline
$(0.85\gev, 0.015\s)$ & $+0.52$ & $(-2.62,+0.42,+2.62)$ & $\simeq0$ & SHiP \\
$(0.6\gev, 0.034\s)$ & $+0.43$ & $(-2.63,+0.38,+2.63)$ & $-0.20$ & SHiP, CMB, GW target \\
$(0.775\gev, 0.02\s)$ & $-0.83$ & $(+1.39,-1.12,+1.79)$ & $+0.02$ & SHiP, C$\nu$B \\
\end{tabular}
\end{ruledtabular}
\end{table*}

{\bf\textit{HNLs, lepton flavor asymmetries, and MeV-temperature Universe.}}--- For $m_{l_\alpha}+m_\pi\lesssim m_N\lesssim1\gev$, the dominant hadronic channels of a Dirac HNL are $N\to\pi^+l^-$ and $\bar N\to\pi^-l^+$. While the injected pions are present, the neutron fraction $X_n=n_n/(n_n+n_p)$ tracks the quasi-static equilibrium of the strong $\pn$ reactions, which overwhelm the weak rates; for a population with number ratio $r\equiv n_N/n_{\bar N}$, it reduces to
\begin{equation}
    X_n\simeq\frac{C_\pi}{C_\pi+r}\,,\qquad C_\pi=\frac{\langle\sigma v\rangle^{\pi^-}_{p\to n}}{\langle\sigma v\rangle^{\pi^+}_{n\to p}}=\mathcal O(1)\,.
    \label{eq:Xn}
\end{equation}
The ratio $C_\pi$ exceeds unity owing to the Coulomb attraction in the $\pi^-p$ channel~\cite{Pospelov:2010cw}. The symmetric case $r=1$ gives the isospin plateau $X_n\simeq1/2$, far above the standard-BBN (SBBN) track; once the pions disappear, the weak interactions are too slow to remove the neutron excess, and helium is overproduced (SuM~\ref{app:mechanism}). The pion-driven conversion alone excludes $\tau_N\gtrsim0.02\s$; the distortion of the neutrino spectra by the decays adds to the helium shift, and our calculation places the bound at $\tau_N\simeq0.012$ to $0.014\s$ across the mass range of Fig.~\ref{fig:money}.

\begin{figure}[t!]
    \centering
    \includegraphics[width=\linewidth]{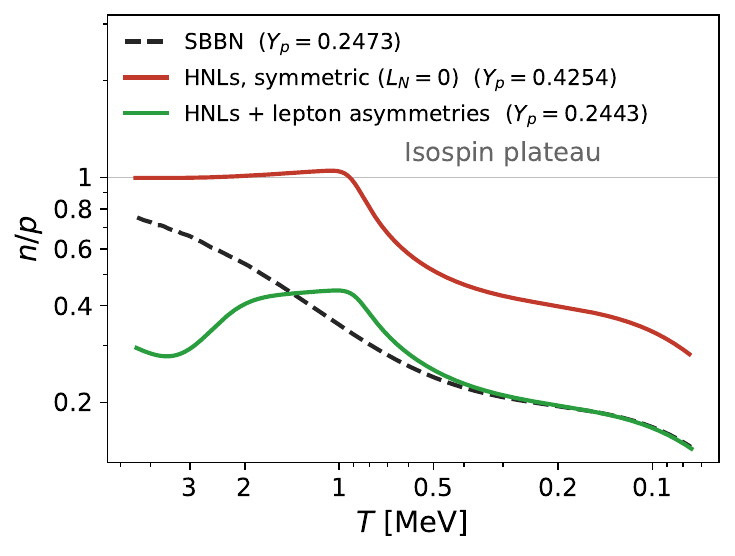}
    \caption{How the BBN bound is avoided in the scenario with HNLs and lepton flavor asymmetries.  A charge-symmetric HNL population (red) drives the $n/p$ ratio to the isospin plateau, far above the SBBN dynamics (black dashed), and helium is overproduced by a factor of $1.7$.  The asymmetric population (green) starts \emph{below} the SBBN track, because a positive electron neutrino degeneracy suppresses $n/p$ by $e^{-\xi_{\nu_e}}$.  The meson-driven conversions then raise it, the two effects compete, and the curve returns to the SBBN line as the mesons disappear.  The lifetime is chosen so that the decays occur at weak freeze-out, where the asymmetry must cancel the meson-driven conversions as they happen.  The point $(m_N,\tau_N)=(0.60\gev,0.069\s)$ is considered for illustration, chosen to lie outside the supernova region of Ref.~\cite{Carenza:2023old}.}
    \label{fig:np}
\end{figure}

Conversely, choosing
\begin{equation}
    r_{\rm req}=C_\pi\,\frac{1-X_n^{\rm SBBN}}{X_n^{\rm SBBN}}\simeq4\text{--}10
    \label{eq:rreq}
\end{equation}
holds the neutron fraction at the value the standard evolution reaches when the mesons die out, and the BBN bounds are relaxed. Fig.~\ref{fig:np} shows the resulting cancellation deep inside the naive exclusion.  Equation~\eqref{eq:rreq} applies while the mesons survive to the $\pn$ freeze-out near $1\s$, which requires $\tau_N\gtrsim0.1\s$; at shorter lifetimes, the degeneracies shift $X_n$ through the weak rates on their own, and $q_N$ compensates that shift, so it may take either sign. The mechanism extends to $m_N\simeq1.5\gev$: well above $1\gev$, $N$ itself emits both $\pi^+$ and $\pi^-$, as well as the charge-blind $K_L$, and the achievable imbalance erodes below $1/r_{\rm req}$ (SuM~\ref{app:mechanism}).

Removing the neutron effect is necessary but not sufficient: the HNLs also modify the neutrino population of the Universe by decaying into non-thermal species, and this modification grows with the abundance $Y_N\equiv(n_N+n_{\bar N})/s$. The decays reheat the electromagnetic plasma relative to neutrinos and \emph{lower} $\neff$~\cite{Sabti:2020yrt,Rasmussen:2021kbf,Akita:2024ork}.

Both conditions, $r\neq1$ and an unchanged $\neff$, may be met in a Universe carrying large lepton flavor asymmetries, $L_\alpha\equiv(n_{\nu_\alpha}-n_{\bar\nu_\alpha}+n_{\alpha^-}-n_{\alpha^+})/s$ with $\alpha=e,\mu,\tau$ and $s$ the entropy density. The chemical potential of the mixing flavor makes $N$ and $\bar N$ unequal, and the degeneracies alone \emph{raise} $\neff$, cancelling the depletion from the decays (SuM~\ref{app:neff}).

{\bf\textit{Dynamics of HNLs and lepton flavor asymmetries.}}--- Here, we consider HNLs that mix with the muon flavor, commonly used as a benchmark model~\cite{Beacham:2019nyx,deBlas:2025gyz}, although the calculation generalizes to arbitrary mixing patterns, with the pattern deciding whether and where the solutions exist (SuM~\ref{app:mixing}).  The setup is specified by the HNL mass $m_N$ and lifetime $\tau_N$ (equivalently the mixing $|U_\mu|^2$), which an accelerator would measure, and by the primordial lepton flavor asymmetries $L_\alpha$.  The asymmetries are defined after sphaleron freeze-out.  At temperatures $T\lesssim10\gev$, where the in-medium suppression of $|U_\mu|^2$ is negligible~\cite{Boyarsky:2020dzc}, $N$ and $\bar N$ come into equilibrium with the plasma, well above the QCD crossover.  This erases any earlier population of these states and transfers part of the asymmetries $L_\alpha$ into the HNL asymmetry $L_N$.  We therefore assume that the HNL population is negligible at $T\gtrsim10\gev$. This assumption does not change the main result, Fig.~\ref{fig:money}, although it may slightly influence the concrete relationship between the HNL parameters and the lepton flavor asymmetries inside the allowed domains (SuM~\ref{app:production}).  Each realization is characterized by the asymmetry stored in the HNL population and by the neutrino degeneracies left in the plasma:
\begin{equation}
    q_N\equiv\frac{L_N}{Y_N}=\frac{r-1}{r+1}\,,\qquad \xi_{20}\equiv(\xi_e,\xi_\mu,\xi_\tau)\Big|_{T=20\mev}\,,
    \label{eq:diagnostics}
\end{equation}
where the degeneracy parameters $\xi_\alpha\equiv\mu_{\nu_\alpha}/T$ are evaluated at the end of the production stage, $T=20\mev$, and carry the evolved flavor asymmetries into the oscillation stage.  Equation~\eqref{eq:rreq} leaves out the modification of the weak $\pn$ conversion by the HNL decays and by the asymmetries: in the full calculation, a large $\xi_e$ shifts the weak equilibrium the mesons act on, and the neutral-current and kaon channels dilute the single-pion balance.

We test the mechanism with a new coupled pipeline of three stages.  The first evaluates the HNL production from $T=30\gev$ down to $T=20\mev$: a momentum-resolved Boltzmann evolution of $N$ and $\bar N$ in a plasma whose conserved primordial charges are redistributed among all species, with the asymmetry drawn into the HNL population and its backreaction tracked exactly (SuM~\ref{app:production}).  From $T\simeq20$ down to $5\mev$, the HNLs begin to decay while neutrino oscillations turn on, and the neutrino and electromagnetic populations can still be described as a single plasma. The second stage computes this evolution with the approach of Refs.~\cite{Domcke:2025lzg,Domcke:2025jiy}, extended by the source from the asymmetric HNL decays.  Below $5\mev$, the third stage evolves the system through neutrino decoupling and BBN with $\nu$DSMC~\cite{Ovchynnikov:2024rfu,Ovchynnikov:2024xyd,Ihnatenko:2025kew,Akita:2026xwu}, extended by the dynamics of metastable decay products of HNLs, following Refs.~\cite{Akita:2024nam,Akita:2024ork}, and follows neutrinos, antineutrinos, and the decaying HNLs together in momentum space. Such a momentum-resolved approach is crucial to trace the neutrino distribution function, which influences both the $\pn$ rates and $\neff$ (SuM~\ref{app:framework} and~\ref{app:validation}).

{\bf\textit{Observational constraints.}}--- The Standard Model prediction for $\neff$ is $\neff^{\rm SBBN}\approx3.044$~\cite{Akita:2020szl,Froustey:2020mcq,Bennett:2020zkv,Ihnatenko:2025kew}, and we define $\dneff\equiv\neff-\neff^{\rm SBBN}$.  The interpretation of the current data is ambiguous: ACT and SPT combined with Planck report $\neff=2.81\pm0.12$~\cite{AtacamaCosmologyTelescope:2025nti,AtacamaCosmologyTelescope:2025blo,SPT-3G:2025bzu}, about two sigma below the Standard Model value, while an analysis of a similar data combination with consistent foreground modelling finds $\neff=3.031\pm0.130\,(\text{stat})\pm0.045\,(\text{fg})$~\cite{Tristram:2025you}.  The Simons Observatory, now operating and aiming at $\sigma_{\neff}\simeq0.045$~\cite{SimonsObservatory:2025wwn}, will test both signs of $\dneff$~\cite{Escudero:2026mgw}.  We adopt $|\dneff|\leq0.05$ as the band in which the HNLs leave no CMB imprint.

We require the primordial nuclear abundances to remain within their observational determinations.  Helium is measured to half a percent, most recently $\yp=0.2458\pm0.0013$~\cite{Aver:2026dxv}, in agreement with the SBBN value $\yp\simeq0.247$~\cite{Pitrou:2020etk}.  The observed deuterium abundance is known to one percent, $10^5\times\text{D/H}=2.55\pm0.03$~\cite{ParticleDataGroup:2024cfk}, but its theoretical prediction is less settled, spanning $2.44\pm0.04$~\cite{Pitrou:2018cgg,Pitrou:2020etk} to $2.51\pm0.07$~\cite{Pisanti:2020efz,Gariazzo:2021iiu} depending on the adopted rates of deuterium burning.  We adopt the higher prediction, which agrees with the data.  The shifts $\Delta\yp$ and $\Delta(\text{D/H})$ that we compute do not depend on this choice; what depends on it is the sign of $\dneff$ that D/H accommodates, because the two shifts are correlated: for $\tau_N\lesssim0.1\s$, we find $\Delta(\text{D/H})/(\text{D/H})\simeq0.11\,\dneff$.  At longer lifetimes, the antineutrinos from the decays convert protons after nucleosynthesis has begun, producing deuterium that does not track $\dneff$, so the coefficient grows (SuM~\ref{app:scan}).  The classification below uses a two-sigma deuterium window, $|\Delta(\mathrm{D/H})|/(\mathrm{D/H})\leq0.031$, matching the two-sigma helium window.

{\bf\textit{Target parameter space.}}--- The helium and $\neff$ conditions fix two combinations of $(L_e,L_\mu,L_\tau)$, and we explore the third direction explicitly.  The HNL mixes with the muon flavor, so its particle-antiparticle ratio $r$ tracks the muon flavor chemical potential at production, and $L_\mu$ is set by requiring that the meson-driven conversion leave $n_n/n_p$ on its standard track when the mesons disappear.  The $\neff$ condition then fixes the size of the neutrino energy excess that compensates the decay heating.  The asymmetry $q_N$ of Eq.~\eqref{eq:diagnostics} and the absolute $L_N$ are then outputs of the production stage.

Deuterium acts in two ways.  For $\tau_N\lesssim0.1\s$, it follows $\dneff$ through the correlation above, where it sets the floor of the admissible $\neff$ window and serves as a consistency check. At longer lifetimes, it no longer does, and the asymmetries reach it only weakly. Two conditions on three asymmetries leave a one-parameter family of solutions at each $(m_N,\tau_N)$: its members differ only in how the net asymmetry is divided between the electron and tau flavors, and $\yp$, D/H, and $\neff$ barely move along it. The windows therefore fix two combinations of $(L_e,L_\mu,L_\tau)$ at the percent level, the muon flavor component and the net asymmetry, and the QCD and C$\nu$B signatures below probe the remaining freedom. Deuterium is never solved for, and beyond $\tau_N$ of a few tenths of a second, no member of the family restores it.

These constraints carve the asymmetry space into a standard $\neff$ domain, in which $\yp$, D/H, and $\neff$ all sit inside the windows above and the HNLs leave no imprint on BBN or the CMB.  This is \emph{the standard $\neff$ region} of Fig.~\ref{fig:money}.

Four physical effects set the boundaries of the domains.  In mass, the charged meson channel closes below $m_\mu+m_\pi\simeq0.245\gev$, and above the GeV scale, multi-hadron final states wash out the correlation between pion charge and lepton number.  In lifetime, the lower boundary is set by heating: near the standard meson bound, the decays deplete $\neff$ by more than the largest admissible asymmetries can compensate.  The upper boundary is set by deuterium, whose excess grows steeply with lifetime.  Every edge in Fig.~\ref{fig:money} ends at a computed solution, at the ceilings of the admissible asymmetries, or at one of these thresholds.  The other cosmological constraints on HNLs arise from decays after nucleosynthesis~\cite{Vincent:2014rja,Diamanti:2013bia,Poulin:2016anj,Domcke:2020ety} and do not reach lifetimes below $1\s$ (SuM~\ref{app:prior-works-comparison}).

Supernovae constrain part of the same region that the upcoming searches will probe~\cite{Carenza:2023old,Chauhan:2023sci}, and the mechanism proposed here does not lift those constraints.  The constraint of Ref.~\cite{Carenza:2023old} covers the domains of Fig.~\ref{fig:money} at $m_N\lesssim0.4\gev$, including the region DUNE and Hyper-K would reach. Near its upper end, the production sits in the Boltzmann-suppressed tail, so the largest mass the constraint reaches responds exponentially to the assumed core conditions (SuM~\ref{app:sn}).  We show it hatched in Fig.~\ref{fig:money}, to mark that it depends on the supernova model.  Even as drawn, the constraint leaves a substantial region within the reach of SHiP.  High-energy neutrino observations from future galactic supernovae~\cite{Akita:2022etk,Fiorillo:2022cdq,Telalovic:2024cot} may test HNLs~\cite{Syvolap:2023trc,Akita:2023iwq} beyond the current supernova bounds.

{\bf\textit{Signatures: from the laboratory to the C$\nu$B.}}--- The lepton asymmetry of the Universe is among the least constrained properties of the Standard Model plasma.  Within our scenario, an HNL discovery at $(m_N,\tau_N)$ inside the domains of Fig.~\ref{fig:money} would constrain the primordial asymmetries to the one-parameter family identified above, along which only the division of the asymmetry between the electron and tau flavors is left free. $\yp$, D/H, and $\neff$ respond to that division too weakly to fix it at discovery-level precision, so it spans a range that closes only with sharper measurements of $\dneff$ and $(m_N,\tau_N)$, which we assume below.  Table~\ref{tab:benchmarks} summarizes the resulting probes.

\emph{The $\neff$ deficit region.}  The residual $\dneff$ is set by how completely the depletion and the compensating rise cancel.  BBN alone allows either sign of the residual, deuterium bounding only its magnitude as quoted above.  We target the negative side for two reasons: it is the direction of the uncompensated HNL effect itself, and it is the direction the measured deuterium abundance accommodates with the prediction adopted above. We take $-0.28\leq\dneff\leq-0.05$ as the window of detectable deficits, its floor set by deuterium through the correlation $\Delta(\text{D/H})/(\text{D/H})\simeq0.11\,\dneff$.\footnote{Repeating the classification with a one-sigma deuterium window, whose floor is $-0.14$, leaves all three domains populated over the entire mass range and inside the SHiP reach, where the standard $\neff$ region is unchanged (SuM~\ref{app:scan}).}  Such deficits will be probed by the Simons Observatory~\cite{SimonsObservatory:2025wwn}, and the scenario joins the models that could account for a preference for $\neff<3$~\cite{Escudero:2026mgw}.

\emph{The GW target region.}  Large flavor asymmetries~\cite{Akita:2025zvq} also drive the chemical potentials of the QCD plasma.  The electron and muon asymmetries act on the quarks through electric neutrality, while the response to $L_\tau$ is suppressed by about $10^{-4}$ at $T\simeq125\mev$, because the charged tau leptons are Boltzmann suppressed.  Once the charge chemical potential exceeds the pion mass, $|\mu_Q|\geq m_\pi$, the confined plasma enters a pion-condensed phase~\cite{Vovchenko:2020crk}, through a first-order transition in a quark-meson model~\cite{Ferreira:2025zeu}.  This phase can leave signatures in the stochastic gravitational wave (GW) background~\cite{Vovchenko:2020crk,Ferreira:2026uzi}.  The GW target region marks the part of the $\neff$ deficit region where $|\mu_Q|$ reaches $m_\pi$ at $T\leq140\mev$, well below the crossover.  The region is a target for dedicated studies: we evaluate the onset of condensation in an ideal hadron-resonance gas and compute neither the transition dynamics nor a GW spectrum.\footnote{The quark chemical potentials, calibrated against Ref.~\cite{Gao:2021nwz}, could also cross the first-order line that starts at the QCD critical endpoint.  However, with the endpoint of current functional QCD calculations~\cite{Gao:2024fhm}, this crossing needs larger asymmetries than the BBN and CMB windows allow (SuM~\ref{app:qcd}).}  Pion condensation also occurs in the standard $\neff$ domain, at all masses and mostly within the SHiP reach (SuM~\ref{app:detailed}).

\emph{The laboratory.} The primary signature is a discovery of an HNL with lifetime $\tau_N\gtrsim0.01\s$, excluded by BBN in the case of zero lepton flavor asymmetries.  Preserving $n_N\neq n_{\bar N}$ over the HNL lifetime requires that $N\leftrightarrow\bar N$ oscillations do not erase the asymmetry before the decays: any Majorana mass splitting must satisfy $\delta m\lesssim\max(\hbar/\tau_N,V_N)$, where $\pm V_N$ are the energy shifts of $N$ and $\bar N$ from forward scattering on the asymmetric plasma.  Across the domains of Fig.~\ref{fig:money}, this gives $\delta m\lesssim10^{-14}$ to $10^{-12}\,$eV (SuM~\ref{app:model}), i.e., lepton number must be conserved to an excellent approximation (becoming an exact symmetry for Dirac HNLs), which is itself a target for searches for lepton number violation.  Accelerator HNL decays are then lepton-number conserving.  The active neutrinos must also be Dirac, and neutrinoless double beta decay is absent: a Majorana mass $m_{\mu\mu}$ generated above $m_N$ would induce $\delta m=|U_\mu|^2|m_{\mu\mu}|$, exceeding that bound for generic values allowed by the oscillation data and by $\Sigma m_\nu<0.12\,$eV.

\emph{Modified relic neutrino sea.}  Even a near-cancellation in $\neff$ need not imply a standard C$\nu$B: the surviving neutrino asymmetries and the decay products leave a $\nu$ sea that is particle-antiparticle asymmetric and carries a hard non-thermal tail.  The relic neutrinos can exceed their standard number by up to about $75\%$, while the antineutrinos are depleted.  The capture rate on tritium, which weights each mass eigenstate by its electron flavor content, rises by up to about $35\%$~\cite{Long:2014zva,Betti:2019ouf}, a long-term target for PTOLEMY-like experiments (SuM~\ref{app:cnub}).  The neutrino excess persists where $|\dneff|$ is tiny, so the relic $\nu$ sea is modified even where BBN and the CMB look standard.  This separates the two ways of evading the meson bound: low-scale reheating~\cite{Hasegawa:2019jsa,Barbieri:2025moq,Akita:2026sqe} removes the HNL population without enhancing the relic $\nu$ sea, so an HNL discovery accompanied by an enhanced sea points to the asymmetry.

The primordial flavor asymmetries are therefore an output of the scenario.  Whether the QCD plasma enters the pion-condensed phase and how strongly the relic $\nu$ sea is enhanced both depend on how the asymmetry is divided between the electron and tau flavors. A gravitational wave detection or a C$\nu$B measurement would settle that division. Depending on $m_N$ and $\tau_N$, moving the electron and tau asymmetries against each other may carry $\dneff$ from the $\neff$ deficit window into the standard $\neff$ window while the pion condensation and the relic sea enhancement both survive. The allowed range of $L_e$ includes negative values.  There, the $\neff$ deficit and the pion condensation hold while the relic sea enhancement is suppressed, and most of the GW target region within the SHiP reach lies at $L_e<0$ (SuM~\ref{app:detailed}).

{\bf\textit{A model realization.}}--- The scenario needs two nontrivial ingredients: large primordial lepton flavor asymmetries, and a Dirac HNL that inherits part of them.  Both arise when right-handed neutrinos with a conserved lepton number extend the Standard Model and an Affleck-Dine condensate generates the asymmetries~\cite{Kawasaki:2002hq,Akita:2025zvq}.  The GeV-mass Dirac HNL is the particle of this setting within laboratory reach, and further right-handed neutrinos give the active neutrinos Dirac masses.  A lighter HNL among them can be the sterile neutrino dark matter, produced resonantly by the asymmetries that BBN and the CMB select. Large lepton flavor asymmetries may bring its mass well above $10\,$keV, which may be crucial given the tension of lighter masses with structure formation observations~\cite{Akita:2025txo}.  The Affleck-Dine condensate also generates the baryon asymmetry~\cite{Kawasaki:2002hq,Akita:2025zvq}.

The setting thus addresses the same three phenomena as the $\nu$MSM~\cite{Asaka:2005pn,Canetti:2012kh}, with other signatures: heavy rather than light sterile neutrino dark matter, Dirac rather than Majorana active neutrinos, and HNLs in the domain naively excluded by BBN.  In the $\nu$MSM itself, the late lepton asymmetries~\cite{Drewes:2026tpy} are about a hundred times smaller than the mechanism of Fig.~\ref{fig:np} needs, and the quasi-Dirac pair has too large a mass splitting (SuM~\ref{app:model}).  If SHiP discovers an HNL and measures $(m_N,\tau_N)$, BBN, the CMB, and a possible $\neff$ deficit fix the asymmetries up to one direction, and the asymmetries predict the X-ray and structure-formation signals of the dark matter as a function of its mass.

{\bf\textit{Conclusions.}}--- We have shown that the meson-driven BBN bound can be lifted for hadronically decaying relics that inherit part of the lepton flavor asymmetries, without modifying their laboratory decay rates.  We propose a new target for accelerator searches: sub-GeV HNLs with lifetimes naively excluded in standard cosmology.  The correlated signatures turn a detection into a measurement of the primordial flavor asymmetries.  An $\neff$ deficit alongside those signatures may point to a decaying relic in SHiP's reach.  Light element abundances near their standard values can conceal a non-standard evolution at MeV temperatures.

\section*{Acknowledgements}
We thank Koichi Hamaguchi for his contributions at the early stages of the project and for carefully reading the manuscript.  We thank Francesco Di Clemente, Benoit Laurent, and Jorinde van de Vis for useful discussions on the QCD transition in the presence of lepton flavor asymmetries, and Miguel Escudero and Maxim Pospelov for carefully reading the manuscript.  The authors used Claude Fable 5 (Anthropic) and GPT~5.6 Sol (OpenAI) for technical assistance.  KA was supported by JSPS KAKENHI Grant Number 24KJ0060.  MO received support from the European Union's Horizon Europe research and innovation programme under the Marie Sk\l odowska-Curie grant agreement No.~101204216.  VS has been partially supported by the Spanish Research Agency through grant \mbox{CNS2023-145338} funded by MCIN/AEI/10.13039/501100011033 and by ``European Union NextGenerationEU/PRTR'', and through Grant \mbox{PID2022-142545NB-C21} funded by MCIN/AEI/10.13039/501100011033/ FEDER, UE and received support from the European Union's Horizon 2020 research and innovation programme under the Marie Sk\l odowska-Curie grant agreement No~\mbox{860881-HIDDeN} and No~\mbox{101086085-ASYMMETRY}, and from the Spanish Research Agency (Agencia Estatal de Investigaci\'on) through the Grant IFT Centro de Excelencia Severo Ochoa No~\mbox{CEX2020-001007-S} funded by MCIN/AEI/10.13039/501100011033, and R\&D\&I Project \mbox{CEX2025-001574-S}, funded by MICIU/AEI/10.13039/501100011033.  The research presented in this publication falls within the research line Particle Physics in the Standard Model and Beyond (BSM).

\bibliography{main.bib}

\clearpage
\onecolumngrid
\twocolumngrid 
\onecolumngrid 

\appendix

\setcounter{equation}{0}
\setcounter{figure}{0}
\setcounter{table}{0}
\setcounter{page}{1}
\makeatletter
\renewcommand{\theequation}{S\arabic{equation}}
\renewcommand{\thefigure}{S\arabic{figure}}
\renewcommand{\thetable}{S\arabic{table}}
\renewcommand{\theHequation}{supp.\Alph{section}.\arabic{equation}}
\renewcommand{\theHfigure}{supp.\Alph{section}.\arabic{figure}}
\renewcommand{\theHtable}{supp.\Alph{section}.\arabic{table}}
\@removefromreset{equation}{section} \@removefromreset{figure}{section} \@removefromreset{table}{section}
\renewcommand{\thepage}{S\arabic{page}}
\setcounter{secnumdepth}{2}
\renewcommand{\thesection}{\Alph{section}}
\makeatother

\begin{center}
\textbf{\large Supplemental Material for\\[0.5ex]
Heavy neutral leptons beyond the BBN bound: probing the lepton asymmetry of the Universe}
\end{center}

\begin{center}
\text{\large Kensuke Akita,\ \ Maksym Ovchynnikov,\ \ Vsevolod Syvolap}
\end{center}

\medskip

The Supplemental Material is organized as follows. Section~\ref{app:mechanism} details the meson-driven neutron dynamics and the HNL mass range in which the mechanism works. Section~\ref{app:framework} describes the three evolution stages and the BBN code: the production stage coupled to the plasma charge redistribution, together with the analytic $\neff$ scaling, then the joint evolution of the decaying HNLs and the neutrino flavor system, the matching between the oscillation and transport stages, and the nuclear network.  Section~\ref{app:scan} gives the solvers and the definitions that locate the standard $\neff$, $\neff$ deficit, and GW target regions.  Section~\ref{app:validation} collects the numerical validation. Section~\ref{app:cnub} defines the cosmic neutrino background observable. Section~\ref{app:model} details the model example and its limitations. Section~\ref{app:sn} discusses the supernova constraints and their model dependence.  Section~\ref{app:qcd} gives the condition for pion condensation, met throughout the GW target region, and the criterion for a first-order QCD transition; both phenomena can leave gravitational wave signatures. Section~\ref{app:detailed} collects the detailed results, including the extended benchmark table and the relaxed-selection maps.

\section{Meson-driven neutron dynamics and the viable mass range}
\label{app:mechanism}

While a population of mesons $h=\pi^\pm,K^\pm,K_L$ is present, the $\pn$ balance is set by the strong meson-driven reactions, which overwhelm the weak interactions,
\begin{align}
    \pi^{-}+p&\to n+\pi^{0}/\gamma, & \pi^{+}+n&\to p+\pi^{0}, \\
    K_L/K^-+p&\to n+X, & K_L/K^-+n&\to p+X,
\end{align}
where $X$ is a multi-pion final state.  The neutron fraction tracks the dynamical equilibrium
\begin{equation}
    X_n(T)=\frac{\sum_h n_h\langle\sigma v\rangle^h_{p\to n}}{\sum_h n_h\big(\langle\sigma v\rangle^h_{p\to n}+\langle\sigma v\rangle^h_{n\to p}\big)}\,,
    \label{eq:Xn-mesons}
\end{equation}
where $n_h$ is the instantaneous meson number density, fixed by the balance between injection from the decays and removal by interactions with nucleons, self-annihilation, and decay. This regime persists until the mesons disappear at a temperature $T_{\rm dis}$, defined by the condition that fewer than one conversion per nucleon occurs below it~\cite{Boyarsky:2020dzc},
\begin{equation}
    \sum_h\int_{t(T_{\rm dis})}^{\infty}\!\!dt\; n_h(T) \left[(1-X_n)\langle\sigma v\rangle^h_{p\to n} +X_n\langle\sigma v\rangle^h_{n\to p}\right]\simeq1\,.
    \label{eq:conversion-number}
\end{equation}
For a symmetric injection the two conversion directions balance, so Eq.~\eqref{eq:Xn-mesons} drives $X_n$ towards the isospin plateau $\mathcal O(1/2)$; once the mesons vanish, the weak interactions are too slow to relax $X_n$ back to the SBBN track, leaving the neutron excess that underlies the near-universal bound $\tau_X\lesssim0.02\s$~\cite{Kawasaki:2017bqm,Fradette:2018hhl,Boyarsky:2020dzc,EscuderoAbenza:2025tsi}.

For the asymmetric Dirac population with $r=n_N/n_{\bar N}$ and a single pionic channel, Eq.~\eqref{eq:Xn-mesons} reduces to Eq.~(1) of the main text, and matching $X_n$ to the SBBN value at $T_{\rm dis}$ gives the required ratio of Eq.~(2). Since $X_n^{\rm SBBN}(T_{\rm dis})\simeq0.13$ to $0.25$ for the lifetimes of interest and $C_\pi=\mathcal O(1)$ (enhanced above unity by the Coulomb attraction in the $\pi^{-}+p$ channel~\cite{Pospelov:2010cw}), one finds $r_{\rm req}\simeq4$ to $10$. Deviations of $X_n$ from the SBBN trajectory at $T>T_{\rm dis}$ are erased by the still-active weak processes; the residual weak rates are themselves modified by the electron neutrino degeneracy, coupling the helium condition to the residual asymmetries (SuM~\ref{app:framework}).  Equation~\eqref{eq:Xn} responds to $r$ with bounded sensitivity, $|\partial\ln X_n/\partial\ln r|=1-X_n<1$ for every $C_\pi$ and $r$, so wherever the conversion dominates the weak rates, the asymmetry moves a quasi-static equilibrium with an order-one response.

\begin{figure}[t!]
    \centering
    \includegraphics[width=0.49\linewidth]{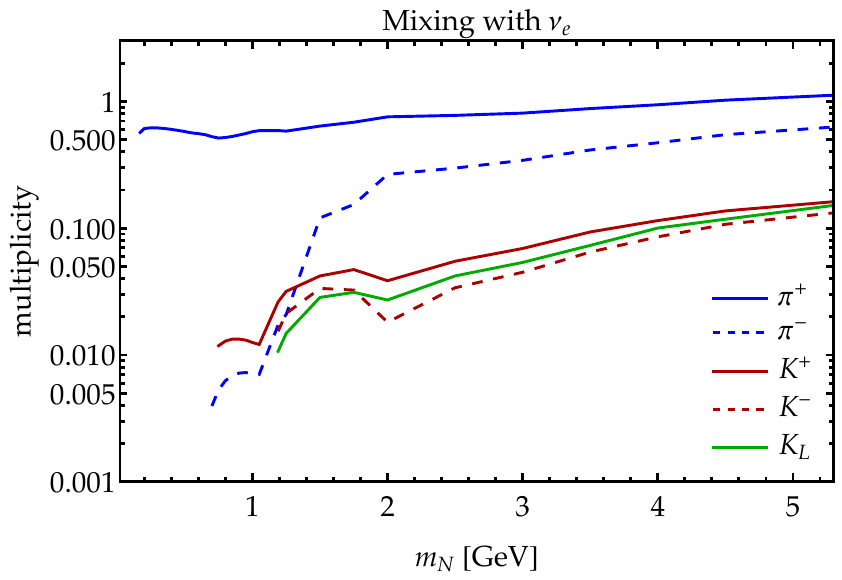}~\includegraphics[width=0.49\linewidth]{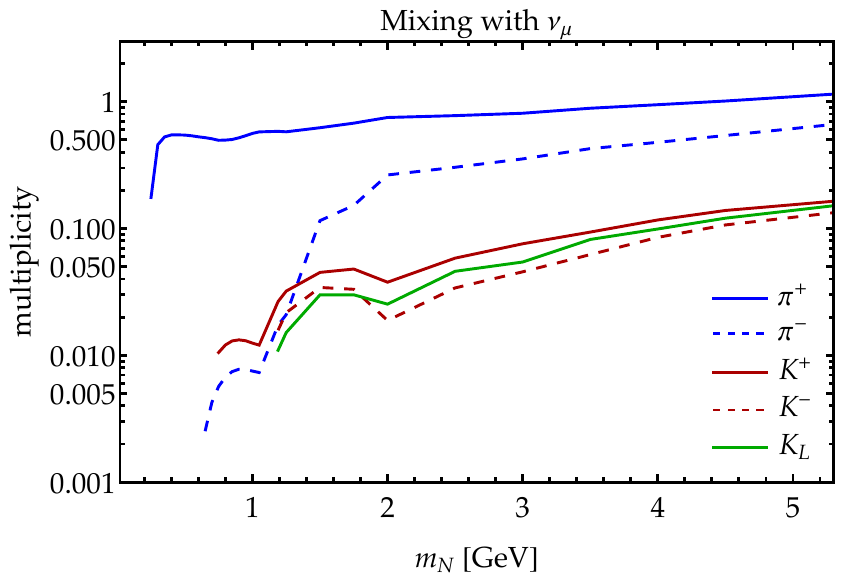}
    \caption{Charged-pion and long-lived-kaon multiplicities per HNL decay versus $m_N$, for mixing with $\nu_e$ (left) and $\nu_\mu$ (right), from combining exclusive decays with the showering of inclusive quark modes~\cite{Ovchynnikov:2023cry}. Below $m_N\simeq1\gev$, $N$ emits essentially only $\pi^+$ (and $\bar N$ only $\pi^-$); at larger masses, $N\to\rho^0\nu,\eta\nu$ and the inclusive $q\bar q'l$/$q\bar q\nu$ channels add $\pi^-$ per $N$ decay at increasing rates, eroding the charge asymmetry.}
    \label{fig:multiplicities}
\end{figure}

\emph{The mass window.} Figure~\ref{fig:multiplicities} shows the mean number of $\pi^\pm$, $K^\pm$, and $K_L$ produced per $N$ decay,
\begin{equation}
 \langle n_h\rangle_N\equiv\sum_c N^c_h\,{\rm Br}(N\to c),
 \label{eq:multiplicity}
\end{equation}
where $N^c_h$ is the number of hadrons of species $h$ in the final state of channel $c$, and analogously per $\bar N$ decay. They are computed with the HNL decay description of Ref.~\cite{Ovchynnikov:2023cry}, in which exclusive channels are used wherever the hadronic final state is resolved, the inclusive quark-level modes are showered and hadronized above their thresholds, and the two are matched in the overlap region.  The total widths, which fix the relation between lifetime and mixing angle, are taken from the same description, which follows the approach of Ref.~\cite{Bondarenko:2018ptm}.  They are Majorana widths, so at a fixed mixing angle a Dirac HNL has half the width and twice the lifetime, and that factor is applied throughout.

The shapes follow from which modes carry the correlation with lepton number.  The charged-current two-body decays $N\to h^+l_\alpha^-$ produce positively charged mesons only, and their conjugates produce negatively charged mesons, so these are the channels that convert the HNL charge asymmetry into a meson charge asymmetry.  The neutral-current modes $N\to h^0\nu$ are blind to it.  Below $m_N\simeq1\gev$, the open hadronic channels are $\pi^+l_\alpha^-$ and $\pi^0\nu$, so $\langle\pi^+\rangle_N$ is of order the charged-current branching fraction while $\langle\pi^-\rangle_N$ stays two orders of magnitude below it, a $\pi^-$ being reachable only through rare multi-body final states.  The per-decay ratio $R_N\equiv\langle\pi^-\rangle_N/\langle\pi^+\rangle_N$ is therefore negligible there, so the charge imbalance per decay is largest.  Above the kaon threshold an analogous hierarchy appears, $K^+$ tracking the charged-current mode while $K^-$ stays rarer, and $K_L$ is charge-blind from the outset, so it dilutes the imbalance without ever contributing to it.  As $m_N$ grows, $N\to\rho^0\nu\to\pi^+\pi^-\nu$ and $N\to\eta\nu$ open, and then the inclusive $q\bar q'l$ and $q\bar q\nu$ modes take over.  All of these produce pions in $\pi^+\pi^-$ pairs, so the two multiplicities rise together and $R_N\to1$.  The matching of exclusive modes onto inclusive showering is a modeling uncertainty near channel thresholds, so no precise mass edge is inferred from a single threshold region. The asymmetric injection can reach the SBBN neutron fraction only if
\begin{equation}
    R_N(m_N)\lesssim\frac{1}{r_{\rm req}}\simeq0.1\text{--}0.25\,,
    \label{eq:window}
\end{equation}
which fails around $m_N\simeq1.5\gev$ in the charged-pion sector alone. In practice, the charge-blind $K_L$ component begins to weaken the mechanism already at $m_N\gtrsim1.2\gev$. Thus the kinematic requirement for a pion-charge imbalance is $m_{l_\alpha}+m_\pi\lesssim m_N\lesssim1.5\gev$. For the muon-mixing figure, the charged-current mechanism begins at $m_\mu+m_\pi\simeq0.245\gev$.  The numerical reach is determined by the transport calculation and need not coincide with this broader kinematic limit.

\emph{The required HNL asymmetry.} The subsequent population is characterized by $Y_N=(n_N+n_{\bar N})/s$ and $L_N=(n_N-n_{\bar N})/s$.  With one populated helicity of each particle and antiparticle, the relativistic $r\simeq5$ estimate is $Y_N\simeq0.55\kappa/g_{*s}^{\rm dec}$ and $L_N\simeq0.365\kappa/g_{*s}^{\rm dec}$, with $L_N/Y_N=(r-1)/(r+1)$ and a possible external entropy dilution $\kappa\leq1$.  The full chain evolves $N$ and $\bar N$ separately with opposite chemical potentials fixed by the conserved plasma-plus-HNL charge, so the asymmetric enhancement of the total density also enters the production era expansion.  The absolute $L_N$ therefore depends on how many HNLs are made and on the entropy the plasma carries while they are present (SuM~\ref{app:production}).

\subsection{Comparison with earlier studies of relaxing the BBN bound}
\label{app:prior-works-comparison}

This work establishes a mechanism by which sub-GeV HNLs with lifetimes $\tau_N\gtrsim0.01\s$ stay consistent with the BBN and CMB observations, over a parameter space that the accelerator-based searches reach in part.  The accelerator phenomenology is unchanged, since the HNL interactions themselves stay \emph{minimal}. The HNLs are produced by their mixing and are present at MeV temperatures.  The resulting cosmology can leave correlated imprints, from the QCD transition epoch near $100\mev$ through the light elements and $\neff$ to the relic neutrino background.

With present precision, the light elements leave little room for departures from standard BBN~\cite{Aver:2026dxv,ParticleDataGroup:2024cfk} and tightly constrain non-standard evolution during the MeV era.  Here they stay inside their observed windows even though the HNL population and the accompanying flavor asymmetries drive that evolution far from standard.  A laboratory measurement of $m_N$ and $\tau_N$ fixes the HNL parameters entering it.  The cosmological effects that normally exclude the particle therefore become probes of the plasma at MeV temperatures.

Earlier studies relaxed the cosmological limits on HNLs by modifying the HNL interactions, by late reheating, or through lepton flavor asymmetries.

\emph{Modified HNL interactions.}  Ref.~\cite{Deppisch:2024izn} considers an extension in which an axion-like particle is present alongside the HNL, so that the HNL decays dominantly into an axion-like particle and a neutrino.  The HNL then disappears earlier and evades the bound, provided the axion-like particles themselves decay considerably later, and a region opens between $1\mev$ and $1\gev$ at mixings $|U|^2$ between $10^{-9}$ and $10^{-6}$.  The mixing angle no longer fixes the laboratory phenomenology, since production still follows $|U|^2$ while the decay rate is set by the new channel, and that reference reports the reach of a decay-in-volume search altered accordingly, while a missing-mass search is untouched.  The total width is set by the new channel, so at a given mixing the HNL is far shorter-lived than the minimal case, and only a small fraction of its decays are visible.  Both act on the upper edge of a beam-dump sensitivity, where the total width enters the exponential that removes HNLs decaying before the detector and the visible fraction multiplies the yield.  The lower edge stays where it was, since there the total width cancels against the visible fraction, so the window is squeezed from above onto a fixed floor.  The axion-like particles themselves decay into neutrino pairs after BBN, which raises $\neff$, and the region that survives lies where that shift stays inside the CMB bound.  That analysis uses a single HNL coupled to one generation, where the ALP decay width to neutrinos is $m_N^2|U|^4m_a/2\pi f_a^2$, carrying the square of the combination that fixes the active neutrino mass in the exact seesaw.  A low-scale seesaw with several HNLs reaches mixings far above that relation through cancellations in the neutrino mass, and the induced $a\nu\nu$ coupling then follows the full flavor structure.  The cosmological windows that depend on when the axion-like particles decay may shift in such a realization.  Ref.~\cite{Dev:2025pru} examines the related proposal that invisible decay modes of the HNL evade the bound.  However, it finds that the presence of such modes \emph{strengthens} the cosmological bounds, through the extra radiation the dark states carry into the BBN epoch and its effect on $\yp$ and $\dneff$.

\emph{Late reheating.}  A low reheating temperature suppresses the HNL population and relaxes the cosmological bounds while leaving the laboratory mixing unchanged~\cite{Gelmini:2004ah,Gelmini:2008fq,Hasegawa:2019jsa,Barbieri:2025moq}. The light element measurements and $\neff$ bound the suppression~\cite{Akita:2026sqe}.  Beyond the $\neff$ deficit, late reheating may deplete the relic neutrino sea~\cite{Akita:2026sqe}, while the present scenario gives a neutrino number excess and the GW target from the QCD transition epoch.

\emph{Lepton flavor asymmetries.}  Closest to the present work, Ref.~\cite{Gelmini:2020ekg} studied HNLs of $150$ to $450\mev$ together with a large lepton asymmetry and found the BBN bounds relaxed.  The HNL considered there is Majorana, a complementary case in which the mechanism studied here does not operate: a Majorana state is its own antiparticle, so its decays emit $\pi^+$ and $\pi^-$ in equal numbers whatever asymmetry the plasma carries, and the charge of the injected pions cannot track it.  The scenario there sits at $\tau_N\gtrsim1\s$, set by a restriction to decay temperatures below $0.5\mev$, where the neutrinos have decoupled from the electromagnetic plasma in standard cosmology.  At those lifetimes, the laboratory reach comes from the kaon-induced flux of DUNE; SHiP covers shorter lifetimes over most of that mass range.  Two points separate that scenario from the mechanism studied here.  First, the mass range it covers lies deep inside the supernova exclusion of Ref.~\cite{Carenza:2023old}, far from its edges, where that constraint is least sensitive to the supernova model (SuM~\ref{app:sn}); the domains opened here reach $1\gev$, beyond the mass at which the exclusion ends.  Second, that analysis omits the meson-driven $\pn$ conversion.  Over the whole mass range it covers, the charged-pion channels are open, and the conversion is the largest effect an HNL has on BBN.  Its lifetimes fall entirely after the thermal weak rates have fallen below the expansion rate, so the injected mesons convert nucleons with nothing left to reverse them. There, the lepton flavor asymmetries act through the neutrino temperature, which is far weaker, so the relaxation reported does not survive once the conversion is included.  At the long end of that lifetime range, the mesons also dissociate nuclei that have already formed, an effect Ref.~\cite{Boyarsky:2020dzc} places at $\tau_N\gtrsim40\s$.

The realizations of the asymmetry also differ.  Ref.~\cite{Gelmini:2020ekg} gives the three flavors one common degeneracy parameter and generates the asymmetry after the HNLs have been produced, so it leaves their abundance untouched.  Here, three independent asymmetries are present before production, and the HNL particle-antiparticle ratio follows from that stage (Sec.~\ref{app:production}).

\section{Evolution framework}
\label{app:framework}

The calculation consists of three evolution stages and the BBN code.  Stage~1, from $T=30\gev$ through the production epoch down to $T=20\mev$, is the production stage below, coupled to the plasma charge redistribution; it computes the HNL abundance, spectrum, particle-antiparticle ratio, and the redistributed residual flavor asymmetries.  Stage~2 evolves the neutrino flavor system in the presence of the decaying HNLs, and follows the collective flavor dynamics from $20\mev$ down to $5\mev$, where stage~3 takes over.  Stage~3, below $5\mev$, follows the momentum distributions of the neutrinos, the decay products, and the metastable species with $\nu$DSMC, and provides the inputs for the BBN code.  Stage~3 pairs the momentum-resolved transport with the momentum-averaged QKE solver of stage~2, integrated on to $0.8\mev$ for the flavor asymmetries, which are imposed on the neutrinos the transport carries (SuM~\ref{app:framework}).  The three stages and the BBN code are described in turn below.  SuM~\ref{app:scan} gives the solvers and the domain definitions, and SuM~\ref{app:validation} the numerical validation.

The three stages below form a single pipeline, each starting from the state its predecessor reaches.  The solvers already exist for simpler settings: the quantum kinetics of large lepton flavor asymmetries with no HNLs present~\cite{Domcke:2025lzg,Domcke:2025jiy}, and the momentum-resolved neutrino transport through decoupling and BBN~\cite{Ovchynnikov:2024rfu,Ovchynnikov:2024xyd,Ihnatenko:2025kew}; Ref.~\cite{Akita:2026xwu} treats decaying relics in that transport at zero lepton asymmetry.  We generalize them here to a large lepton flavor asymmetry in the presence of HNLs: the flavor system is sourced by the asymmetric decays, the transport is driven toward the asymmetries that system computes, and the HNL particle-antiparticle ratio follows from the production stage.  Momentum-resolved treatments of the neutrino sector exist for the standard case~\cite{Froustey:2020mcq,Akita:2020szl,Bennett:2020zkv,Ihnatenko:2025kew} and for large primordial asymmetries~\cite{Froustey:2021azz,Froustey:2024mgf}; to our knowledge none of them carries a relic that injects non-thermal neutrinos.  Ref.~\cite{Gelmini:2020ekg}, the only earlier study of HNLs with large lepton asymmetries, omits both the meson-driven conversion that dominates this mass range and the momentum-resolved neutrino evolution (SuM~\ref{app:mechanism}).

This is not an exact solution of the coupled problem, and two of its approximations are structural.  The flavor sector is momentum averaged while the transport is momentum resolved, so the transport has to place the asymmetries at energies that the flavor sector does not determine; the rule it uses is given with stage~3, and its size is measured in SuM~\ref{app:validation}.  Stages~2 and~3 run adiabatically, with the asymmetry-dependent $\nu$--$\nu$ potential switched off.  Each stage makes further simplifications, stated where that stage is described, and SuM~\ref{app:validation} and SuM~\ref{app:convergence} measure how far the observables move under the principal modeling choices and numerical settings.  The calculation therefore establishes that the solutions exist and which signatures they correlate, and locates the domain boundaries to the accuracy measured there.

Parameter counting settles only the dimension of a solution set once that set is known to be non-empty.  Wherever a regular solution exists and the two observational conditions on the three primordial asymmetries are independent, the set is a one-parameter family of asymmetries.  Existence is a dynamical result.  The map from $(L_e,L_\mu,L_\tau)$ to $\yp$ and $\neff$ is fixed by the coupled evolution, and HNL production and decay, the meson-driven $\pn$ conversion, the flavor redistribution, and the thermalization of the non-thermal decay neutrinos could correlate the two observables so that no asymmetry restores the light element abundances and compensates the shift in $\neff$ at the same time.  The evolution calculation shows that they do not, over an extended range of $(m_N,\tau_N)$.

The calculation therefore supports two claims of different strength.  That the compensating solutions exist, and which signatures they correlate, follows from the evolution as a whole.  Which primordial asymmetries a given $(m_N,\tau_N)$ requires, and with them the exact edges of the domains, depends in addition on the flavor treatment.  A fully momentum-dependent quantum-kinetic treatment would describe the flavor evolution more completely, and it could move the asymmetries that a given laboratory point requires, and with them the boundaries and the correlated signatures.  The framework here is built to settle existence and the qualitative structure first: it keeps the quantum-kinetic evolution of the integrated asymmetries and joins it to the momentum-resolved thermalization of the non-thermal neutrinos, the two pieces of physics that a complete treatment would unify.  Such a unified calculation would sharpen the map between $(m_N,\tau_N)$ and $(L_e,L_\mu,L_\tau)$, by an amount that remains to be determined.

\subsection{Stage 1: HNL production}
\label{app:production}

This appendix specifies the production stage, which runs from $T=30\gev$ down to $20\mev$ and computes the initial conditions for the subsequent oscillation and transport stages (SuM~\ref{app:framework}).  We distinguish the robust qualitative statement that mixing produces HNLs while they are relativistic from the stronger statement that they reach a complete equilibrium abundance.

\emph{Production channels and the production epoch.}  HNLs may be produced resonantly, where the in-medium mixing angle $U_{\rm m}$ is enhanced by the lepton flavor asymmetries entering the matter potential $V_{\rm asy}\simeq\sqrt2G_FL_\alpha s$~\cite{Shi:1998km}, and non-resonantly, at the rate $\Gamma\simeq c_\alpha G_F^2pT^4\,U_{\rm m}^2$, with $c_\alpha$ the flavor-dependent numerical coefficient of the collision integrals.  We first show that the resonant channel may be neglected.  The asymmetric potential crosses $\Delta=m_N^2/(2p)$ at
\begin{equation}
 T_{\rm res}\simeq7.3\gev \left(\frac{80}{g_*}\right)^{1/4} \left(\frac{0.1}{|L_\alpha|}\right)^{1/4} \left(\frac{m_N}{\gev}\right)^{1/2} \left(\frac{3}{p/T}\right)^{1/4}.
\end{equation}
The scale is momentum-dependent through $\Delta\propto1/p$; we quote it at the thermal-average momentum $p\simeq3T$.  This crossing lies inside the domain where the HNLs of interest are in ordinary equilibrium, $\Gamma\gg H$, with $H$ the Hubble rate.  We verified this at the point of Fig.~\ref{fig:money} with the smallest mixing angle, the lower-right corner of the colored domains, $(m_N,\tau_N)\simeq(0.95\gev,0.17\s)$: there the computed rate exceeds $H$ from $T\simeq12\gev$ down to $T\simeq2\gev$, and everywhere else in the domains the mixing angle is larger.  Between $T\simeq12$ and $2\gev$, any departure of the HNL distributions from equilibrium decays as $e^{-\int\Gamma\,dt}$, with $\int\Gamma\,dt$ of several hundred for $N$ and about ten for $\bar N$: the asymmetric matter potential enhances the effective mixing of $N$ when it crosses its resonance, while $\bar N$ has no resonance to cross.  Any population the crossing produces relaxes to the Fermi-Dirac attractors of Eq.~\eqref{eq:production-boltzmann}, so the abundance and the asymmetry at $20\mev$ do not depend on it, and the crossing needs no momentum-space resolution.  The non-resonant channel contains no enhancement, and its epoch follows from the in-medium suppression of the mixing angle.  The CP-even thermal potential, $|V_T|\simeq a_\alpha G_F^2pT^4$ with the analogous coefficient $a_\alpha=\mathcal O(10)$, suppresses the mixing as
\begin{equation}
 U_{\rm m}^2\simeq\frac{U_\alpha^2}{\left(1+|V_T|/\Delta\right)^2},
\end{equation}
so the ratio of the rate to $H$ scales as $\Gamma/H\propto T^5U_{\rm m}^2/T^2=T^3U_{\rm m}^2$ at $p\simeq3T$.  It grows as $T^3$ while $U_{\rm m}\simeq U_\alpha$ and falls as $T^{-9}$ once $|V_T|\gtrsim\Delta$, so the maximum lies at $|V_T|\simeq\Delta$.  For vanishing asymmetries, this gives
\begin{equation}
 T_{\rm max}\simeq27\gev \left(\frac{m_N}{\gev}\right)^{1/3} \left(\frac{3}{p/T}\right)^{1/3}.
\end{equation}
For the large asymmetries considered here the total potential is dominated by $V_{\rm asy}$ below $T\sim100\gev$, and the maximum of $\Gamma/H$ moves to the crossing $T_{\rm res}$.  At the point with the smallest mixing angle, the computed rate for $N$ peaks at $T_{\rm res}$, near $6\gev$.  In either case, production is concentrated at temperatures of a few to tens of GeV.  This has two consequences used throughout: the HNLs are produced relativistic, since these temperatures far exceed the sub-GeV masses, and production occurs after sphaleron freeze-out, so the primordial asymmetries are already fixed when the HNL population builds up.  These estimates locate the production epoch; every quoted yield and asymmetry follows from the kinetic equation below.

\emph{Kinetic equation and state counting.}  We use a momentum-resolved relaxation-time Boltzmann equation with the interaction rates $\Gamma_\pm(p,T;\boldsymbol\mu,m_N)$ of $N$ ($+$) and $\bar N$ ($-$), computed from the finite-temperature collision integrals of Refs.~\cite{Akita:2022hlx,Akita:2025txo}, augmented by the asymmetric potential of Ref.~\cite{Shi:1998km}.  The rates are evaluated separately for the two charge states on the plasma state of the five-potential charge redistribution below, so they carry the full linear dependence on the lepton- and quark-sector chemical potentials;  the external HNL leg enters the collision kinematics with its massive dispersion, $E=\sqrt{p^2+m_N^2}$.  The CP-even thermal potential and the collision damping regulate the asymmetric resonance.  The asymmetries therefore act on the production in three places: in the rates $\Gamma_\pm$ themselves, in $V_{\rm asy}$, which enters the in-medium suppression with opposite signs for $N$ and $\bar N$, and in the Fermi-Dirac attractors at $\mp\xi_\mu$ of Eq.~\eqref{eq:production-boltzmann}; the quark chemical potentials contribute both through the charge-neutrality response of the plasma and directly through the quark-channel rates.  The residual rate-level approximations are the contact form of the weak vertices, appropriate at the production temperatures reached here, and the momentum-averaged rescaling below the QCD crossover, described at the end of this appendix.  Mixing populates one helicity state each of the Dirac particle $N$ and of its antiparticle $\bar N$.  The implementation evolves their momentum distributions separately,
\begin{equation}
 \left(\partial_t-Hp\partial_p\right)f_\pm= \Gamma_\pm(p,T;\boldsymbol\mu,m_N) \left[f_{\rm FD}\!\left(E/T\mp\xi_\mu\right)-f_\pm\right] -\frac{m_N}{E\,\tau_N} f_\pm ,
 \label{eq:production-boltzmann}
\end{equation}
with the signs $+$ and $-$ denoting $N$ and $\bar N$.  The first term describes the HNL production in $2\to2$ reactions with the plasma, and the second term the HNL decay.  For the lifetimes of interest, the decays become efficient only at temperatures well below $m_N$, where the inverse decays are Boltzmann-suppressed and are therefore omitted.  When reactions are rapid, the two distributions approach opposite chemical potentials, $\mu_N=\mu_{\nu_\mu}$ and $\mu_{\bar N}=-\mu_{\nu_\mu}$; away from that limit, the abundance and the asymmetry are both outputs of Eq.~\eqref{eq:production-boltzmann}.

\emph{Plasma charge redistribution.}  At every step of Eq.~\eqref{eq:production-boltzmann} the plasma state follows from a simultaneous solution for the five chemical potentials below: every species $i$ carries $\mu_i=B_i\mu_B+Q_i\mu_Q+\sum_\alpha L_{\alpha,i}\,\mu_{L_\alpha}$, and the five potentials are fixed by the three comoving flavor asymmetries, the baryon charge, and electric neutrality, with exact Fermi-Dirac and Bose-Einstein integrals.  Above $250\mev$, the gas comprises photons, gluons, charged leptons, neutrinos, and the $u,d,s,c,b$ quarks; below $160\mev$, a truncated hadron-resonance gas (with a check against Bose condensation of the charged mesons) sets the asymmetry corrections, blended across the crossover window as relative corrections to the tabulated $g_{*s}$ baseline.  Two consequences are important.  First, the conserved charges are distributed over all plasma species, so the degeneracy of the mixing flavor, $\xi_\mu=\mu_{\nu_\mu}/T$, comes out smaller than the flavor charge alone would suggest.  Since $\xi_\mu$ sets the $N$--$\bar N$ imbalance, that difference carries into $r=n_N/n_{\bar N}$.  Second, the charge-neutrality response couples the flavors: large electron- or tau-flavor asymmetries raise $\mu_{\nu_\mu}$ at the production epoch, so $r$ depends on all three asymmetries and the muon flavor component required by the helium condition of the main text can be small and negative.

\emph{Charge conservation, expansion, and entropy.}  The lepton asymmetry is divided between the plasma and the HNL population, and their sum is conserved,
\begin{equation}
 L_{\rm tot}=L_{\rm plasma}+L_N, \qquad L_N=Y_N^{(+)}-Y_N^{(-)}, \qquad Y_N=Y_N^{(+)}+Y_N^{(-)},
 \label{eq:production-charge}
\end{equation}
which the calculation follows as an exact comoving balance, with a midpoint correction for the plasma-HNL charge exchange within a step; its residual effect on $r$ is below $10^{-4}$.  The ledger of Eq.~\eqref{eq:production-charge} and the output abundance both count the two helicity states of each species, $g_N=2$; this counting generates Fig.~\ref{fig:money} and the values of Tables~\ref{tab:benchmarks} and~\ref{tab:benchmarks-extended}.  The Hubble rate includes the HNL energy density and the degenerate excess of the full plasma, $\Delta\rho(\boldsymbol\mu)=\rho_{\rm gas}(\boldsymbol\mu)-\rho_{\rm gas}(0)$, and the scale factor follows from imposing conservation of the total comoving entropy, $s_{\rm SM}(T,\boldsymbol\mu)\,a^3+S_N^{\rm c}={\rm const}$, with $s_{\rm SM}(T,\boldsymbol\mu)$ the Standard Model entropy density evaluated at the chemical potentials of the charge redistribution, where $S_N^{\rm c}$ is the exact kinetic entropy of the evolving $N/\bar N$ distributions, $s_N=-g\!\int\!\frac{d^3p}{(2\pi)^3}\left[f\ln f+(1-f)\ln(1-f)\right]$, valid also away from equilibrium.  The entropy generated by the collisions themselves is omitted from this balance.  It is second order in the departure from equilibrium, which is small while the HNLs are held at their attractors, and the HNL population carries at most $2\%$ of the comoving entropy, so the error it leaves on the dilution history is well below that.  \par\emph{Initial HNL population and helicity states.}  The production stage starts at $T=30\gev$, the highest temperature at which the rates $\Gamma_\pm$ are computed, with no HNLs present, $f_\pm=0$.  The values of $Y_N$ and $r$ at $20\mev$ are insensitive to this choice.  We repeated the calculation, with the same conserved charges, from four other initial populations: charge-symmetric equilibrium, equilibrium at the plasma chemical potentials, twice the equilibrium abundance, and an HNL asymmetry of the opposite sign.  Each of them changes $Y_N$ and $r$ by less than $0.06\%$ at $(0.7\gev,0.022\s)$ and $(0.6\gev,0.14\s)$, and by less than $1\%$ at the point of Fig.~\ref{fig:money} with the smallest mixing angle, $(m_N,\tau_N)\simeq(0.95\gev,0.17\s)$.  HNLs from decays of on-shell $W$ bosons, which can dominate freeze-in production above $m_N\simeq0.23\gev$~\cite{Li:2026kct}, join this early population and are erased in the same way.  All four populations consist of the helicity states that mixing produces and Eq.~\eqref{eq:production-boltzmann} evolves.\par The Yukawa coupling of the HNL to the Higgs and lepton doublets creates the opposite helicity states, positive for $N$ and negative for $\bar N$, and the production stage assumes that they follow the distributions $f_\pm$.  After electroweak symmetry breaking, these states couple to the plasma through helicity flips suppressed by about $(m_N/E)^2$ and, near the electroweak scale, through inverse decays of the $W$, $Z$, and Higgs bosons.  Evolving them with both processes, from initial abundances between zero and charge-symmetric equilibrium, changes $Y_N$ by $5$ to $20\%$ and $r$ by $5$ to $35\%$ at the same three points.  If the $\bar N$ states of negative helicity are never populated, $r$ stays within a factor of $1.8$ of the value that the production stage gives.  The ratio $r$ depends steeply on the muon asymmetry, $d\ln r/dL_\mu\simeq60$, so solving the helium condition again for $L_\mu$ compensates these shifts with $|\Delta L_\mu|\lesssim0.01$, no larger than the spacing between adjacent computed members of the family of asymmetries.  The accompanying change in the energy that HNL decays inject moves $\dneff$ by at most a few hundredths, and since $d(\dneff)/dL_e\simeq2$ to $5$, a correction $|\delta L_e|\lesssim0.02$ offsets it.

The production stays on down to $20\mev$.  Below the QCD crossover, the collision term keeps the momentum dependence and the charge split of its $250\mev$ form, and the magnitude follows the total interaction rate of a massless $\nu_\mu$, including the hadronic-phase channels of Ref.~\cite{Akita:2025txo}, relative to its value there; this factor reaches $0.18$ at $20\mev$.  For $m_N>0.4\gev$, the mixing angle that the mass and the lifetime fix, $|U_\mu|^2=f(m_N)/\tau_N$ with $f$ steeply decreasing, remains $\lesssim2\times10^{-8}$ across the domains of Fig.~\ref{fig:money}, so $\Gamma/H\propto T^3|U_\mu|^2$ falls to unity already at $T\simeq0.25$ to $2\gev$, before the transition.  No efficient channel opens below the crossover either: the decay $\pi\to\mu N$ is kinematically closed, kaon decays $K\to\mu N$ add at most $5\times10^{-4}$ of the equilibrium abundance, inverse decays at most $2\%$, and the leptonic $2\to2$ reactions and the scatterings on mesons together give about half the collision rate just above it.  This late production contributes at most $7\%$ of the yield for $m_N\geq0.4\gev$, and up to $60\%$ at $m_N=0.25$ to $0.4\gev$.  Halving or doubling the confined-phase factor shifts $\dneff$ there by about $0.01$, never beyond $0.07$: every region keeps these masses, and no edge of the drawn domains moves by more than one step of the lifetime grid.  The stage returns the abundance $Y_N$, the ratio $r$, the momentum spectrum, and the redistributed residual flavor asymmetries at $20\mev$, which initialize the evolution that follows.

Once production has stopped and no lepton-number-violating interaction acts, the surviving comoving HNL asymmetry is conserved until the HNLs decay.

\subsection{Analytic estimate of the $\neff$ balance}
\label{app:neff}

The useful analytic intuition is an energy-budget cancellation: the non-relativistic HNL energy present near decay tends to heat the electromagnetic plasma relative to neutrinos, whereas an active-sector degeneracy raises the neutrino energy density.  For orientation only, a radiation-dominated standard-BBN time-temperature relation gives
\begin{equation}
 T(\tau_N)\simeq1\mev \left(\frac{0.738\s}{\tau_N}\right)^{1/2} \left(\frac{10.75}{g_*}\right)^{1/4}.
 \label{eq:analytic-decay-temperature}
\end{equation}
The full calculation integrates the time-temperature relation directly, which allows Eq.~\eqref{eq:analytic-decay-temperature} to be tested against it: at $t=\tau_N$, the standard-BBN chain reproduces Eq.~\eqref{eq:analytic-decay-temperature} to $0.1\%$, and the benchmark of Fig.~\ref{fig:asym-evolution} to $6\%$, the deviation being the backreaction of the degenerate plasma and of the partly non-relativistic HNLs on $H(T)$.

For $m_N\gg T$, $\rho_N+\rho_{\bar N}\simeq m_NY_Ns(T)$.  In the illustrative common-degeneracy ansatz, the excess energy of three active flavors is
\begin{equation}
 \frac{\Delta\rho_\nu}{\rho_{\nu,\rm tot}(0)} =f(\xi_\nu)=\frac{30}{7}\left(\frac{\xi_\nu}{\pi}\right)^2 +\frac{15}{7}\left(\frac{\xi_\nu}{\pi}\right)^4,
\end{equation}
so the scaling $\rho_N+\rho_{\bar N}\sim\Delta\rho_\nu$ gives
\begin{equation}
 f(\xi_\nu)\simeq\frac{16}{63}g_{*s}(T) \frac{m_NY_N}{T(\tau_N)}.
 \label{eq:analytic-neff-balance}
\end{equation}
Written for a common degeneracy, the left-hand side generalizes to a flavored state as $\frac13\sum_\alpha f(\xi_\alpha)$, which is how we evaluate it below.  Equation~\eqref{eq:analytic-neff-balance} is then a calibrated upper envelope: over the $\neff$-deficit domain (SuM~\ref{app:scan}) its right-hand side tracks the computed degeneracies with a correlation of $0.91$ in the logarithm and overestimates them by a factor $2.6$, with a $16$th to $84$th percentile spread of $2.3$ to $3.0$.  The scaling with $m_NY_N/T(\tau_N)$ is therefore correct, and the overshoot is the energy that the pion-muon chains return to the neutrinos, which the estimate ignores; the large ratio $m_N/T$ makes degeneracies of order unity necessary at all.  Given $(m_N,\tau_N)$ and the production yield, dividing Eq.~\eqref{eq:analytic-neff-balance} by that factor predicts the required degeneracy to within tens of percent; the coupled production, oscillation, and transport calculation determines every quoted value.

\subsection{Stage 2: neutrino oscillations in the presence of HNLs}

Stage~2 is initialized from the output of the production stage: the initial degeneracies follow from the residual redistributed flavor asymmetries at $20\mev$, and the decay source uses the same population's abundance, spectrum, and $N/\bar N$ asymmetry.

In our scenario, the evolution of the plasma is affected both by the large lepton flavor asymmetries and by the HNLs.  The HNLs store energy density and release it in their decays, which also inject lepton flavor asymmetries, and the resulting problem is the dynamics of the neutrino oscillations that redistribute them.  We therefore use the approach of Refs.~\cite{Domcke:2025lzg,Domcke:2025jiy}, a quantum-kinetic framework for the evolution of the Universe in the presence of lepton flavor asymmetries (COFLASY), and incorporate the HNLs in it as both a component of the energy density and a source of asymmetries.  The evolved quantities are momentum-averaged $3\times3$ neutrino and antineutrino density matrices, represented by 18 Gell-Mann coefficients, together with the ratios of the neutrino and photon temperatures to the comoving reference temperature $T_{\rm cm}\propto1/a$, namely $T_\nu/T_{\rm cm}$ and $T_\gamma/T_{\rm cm}$.  We use the adiabatic approximation of those references.

The coupling to the decaying HNLs is operator split: within each step of $d\ln a$, the two sectors are advanced in sequence.  First the energy density $\rho_N$ and the pressure $P_N$ of the HNL population are computed from its momentum distribution and enter the expansion,
\begin{equation}
 H^2=\frac{8\pi G}{3}\left[\rho_{\rm SM}(\rho,\bar\rho,T_\nu,T_\gamma)+\rho_N+\rho_{\bar N}\right],
 \label{eq:qke-friedmann}
\end{equation}
and within the step this component is redshifted with the local equation of state $P_N/\rho_N$.  The quantum kinetic equations of Refs.~\cite{Domcke:2025lzg,Domcke:2025jiy},
\begin{equation}
 i\frac{d\rho}{dt}=\left[\mathcal{H},\rho\right]+\mathcal{C}[\rho],\qquad
 i\frac{d\bar\rho}{dt}=\left[\bar{\mathcal{H}},\bar\rho\right]+\mathcal{C}[\bar\rho],
 \label{eq:qke}
\end{equation}
are then integrated across the step at fixed HNL density, advancing the oscillations, the collisions, and the two temperatures.  Here $\mathcal{H}$ holds the vacuum term and the matter potential, whose asymmetry-dependent part enters with opposite signs in $\mathcal{H}$ and $\bar{\mathcal{H}}$, and $\mathcal{C}$ is the collision term.  The decays are applied at the end of the step as an injection into the six neutrino species and into the electromagnetic plasma,
\begin{equation}
 n_i\to n_i+\Delta n_i,\qquad \rho_i\to\rho_i+\Delta\rho_i,\qquad \rho_{\rm EM}\to\rho_{\rm EM}+\Delta\rho_{\rm EM},
 \label{eq:qke-injection}
\end{equation}
with $i$ running over $\nu_e,\nu_\mu,\nu_\tau$ and their antiparticles.  Integrating $H$ gives the elapsed physical time, which sets the decayed fraction of the HNL population.  The simulated decays of $N$ and $\bar N$, resolved by the electric charge of every final state, together with the fate of each injected meson (absorption on nucleons, decay, or annihilation), determine the number and energy injected into each neutrino and antineutrino species and the energy deposited in the electromagnetic plasma.  The updated radiation densities affect the next Hubble step.  Thus the feedback through $H$ and elapsed time is physical, but the two sectors are advanced in sequence within each step, so the splitting error is of first order in the step size, and the evolution contains neither HNL inverse reactions nor a momentum-dependent collision operator coupling the HNLs to the neutrinos.

The density matrices are flavor resolved, so the asymmetries are followed flavor by flavor throughout.  The energies are not: COFLASY carries one neutrino temperature common to all six species.  Each injection preserves the six species number increments, the three net flavor asymmetries, and the total injected neutrino energy, while the difference between the flavor-resolved energy injected by the simulated decays and the energy change the single-temperature state realizes in that flavor is discarded.  That difference cancels in the sum over the six species, and it measures how the injected energy is divided among the flavors, which the collisions themselves drive while the species remain coupled.  Its impact on the observables is bounded in SuM~\ref{app:validation}.  The neutron fraction $X_n$ is evolved along the same physical time, with weak rates evaluated on the instantaneous electron neutrino number and energy densities and with the same meson-driven conversion rates.  During this stage non-neutrino decay energy is deposited electromagnetically, and the charged-lepton contribution to the lepton charge is carried unchanged.

\subsection{Stage 3: momentum-resolved neutrino evolution}

Neutrinos begin to decouple around $T\simeq5\mev$, and from there on their spectra have to be followed in momentum, where the earlier stages needed only integrated densities.  Resolving the momentum evolution matters for both $\neff$ and the $\pn$ conversion rates~\cite{Boyarsky:2021yoh,Ovchynnikov:2024xyd,Akita:2024nam}, and it sets the upper lifetime edge of the domains.  The decays leave a non-thermal antineutrino population that only redshifts, and oscillations keep converting it into the electron flavor whatever flavor each antineutrino was produced in.  Those above the $\bar\nu_e\,p\to n\,e^+$ threshold create neutrons after deuterium burning has ended.  How many stay above it depends on the shape of the spectrum, which the energy density alone does not fix.  An exact treatment would evolve the full quantum kinetic equations at every momentum, which is prohibitively slow for a parameter scan.  A momentum-resolved Boltzmann transport is fast enough and can carry neutrinos and antineutrinos separately, but it does not by itself reproduce the quantum-kinetic flavor redistribution of the large asymmetries.  We therefore use a hybrid of two calculations.  The momentum-resolved thermalization is solved with the Boltzmann approach of $\nu$DSMC~\cite{Ovchynnikov:2024rfu,Ovchynnikov:2024xyd,Ihnatenko:2025kew}; we call this the transport.  The redistribution of the integrated asymmetries is computed by the momentum-averaged quantum kinetic solver of stage~2, integrated on below $5\mev$; we call this the QKE solver.  Its asymmetries are imposed on the neutrinos the transport carries.  The $\nu$DSMC treatment of decaying relics is presented in Ref.~\cite{Akita:2026xwu}.

$\nu$DSMC represents the six neutrino species by a sample of Monte-Carlo particles in a comoving volume, each carrying an energy and a propagation-state label.  Collisions are drawn pairwise from that sample, and the occupancies entering the Pauli-blocking factors are estimated from it as well.  The six species are the three neutrino and three antineutrino flavors, counted separately.  The comoving volume is set so that a species at the mean of the six number densities carries the chosen number of Monte-Carlo neutrinos; every species then carries a number proportional to its own density, so a species at a larger degeneracy carries proportionally more, and the pool totals six times the chosen number.  That number fixes the statistical precision.  Every quoted result uses $3.2\times10^4$ per species on average, about $1.9\times10^5$ in the whole pool; the convergence with it is examined in SuM~\ref{app:convergence}.

For the HNLs, the results are sensitive to the exponentially small population that survives to late times, and a sample of particles would be left with too few of them to resolve that tail.  They are therefore carried on a comoving momentum grid, with the spectrum inherited from the production stage, each bin decaying at the rate its own time dilation implies.  The decays are drawn from that grid, with $\pi^+$ and $\pi^-$ counted separately, and their products enter the Monte-Carlo pool and the metastable population, where the injected mesons and muons are evolved explicitly~\cite{Akita:2024nam,Akita:2024ork}; the resulting neutrino spectra set the momentum-resolved weak $\pn$ rates.

At $5\mev$, stage~3 starts from the elapsed physical time, the surviving HNL distribution, the number and energy densities of the six neutrino species, and the flavor asymmetries of stage~2.  The asymmetries keep evolving below that point: HNLs at the long-lifetime edge of the domains are still decaying there, and stage~3 lets those decays deposit their charge directly.  The neutrinos are also not yet fully decoupled, least of all the energetic neutrinos from the decays, whose cross sections grow with energy, so stage~3 resolves their collisions in momentum.

The transport initializes each neutrino species with a Fermi-Dirac-shaped distribution that matches its number and energy densities.  This keeps the number and energy density of each species and discards the flavor coherences, which stage~3 does not carry; relative to the diagonal rates they stay below $10^{-5}$ over the range where stage~3 runs.  No spectrum is lost with them, because the QKE solver carries none: it evolves momentum-averaged density matrices at a common neutrino temperature, so its spectra are thermal by construction, and the distortions are generated in stage~3 itself.

\emph{Oscillations.}  Each Monte-Carlo neutrino carries a propagation label $i$, assigned when the particle is created and kept until it interacts.  A neutrino produced in flavor $\alpha$ at energy $E$ receives label $i$ with probability $|U^{\rm m}_{\alpha i}(E,T)|^2$, where $U^{\rm m}$ diagonalizes the vacuum Hamiltonian plus the CP-even thermal charged-lepton potential.  Collisions are evaluated in that basis.  For an incoming state $i$ at energy $E$ and an outgoing state $j$ at $E'$, the neutral-current and charged-current pieces combine into
\begin{equation}
 C^{L}_{ji}=g_L^{x}\,O_{ji}+U^{{\rm m}*}_{ej}(E')\,U^{\rm m}_{ei}(E),
 \qquad
 C^{R}_{ji}=g_R\,O_{ji},
 \label{eq:prop-couplings}
\end{equation}
with the overlap
\begin{equation}
 O_{ji}=\sum_\alpha U^{{\rm m}*}_{\alpha j}(E',T)\,U^{\rm m}_{\alpha i}(E,T).
 \label{eq:prop-overlap}
\end{equation}
Here $g_L^{x}=-1/2+\sin^2\theta_W$ and $g_R=\sin^2\theta_W$ are the neutral-current couplings of the muon and tau flavors.  The neutral current treats the three flavors alike and enters through $O_{ji}$ alone.  The charged current adds unity on the electron projection, raising the left coupling of that flavor to $g_L^{e}=+1/2+\sin^2\theta_W$, which separates the electron rate from the muon and tau rates.  Labels change only through the amplitudes of accepted collisions, so the forward and reverse kernels are adjoint by construction.  Pair creation inserts its daughters by the same rule.  The off-diagonal elements of the transported density matrix are dropped.

The QKE solver and the transport both run in the adiabatic approximation, in which the asymmetry-dependent $\nu$--$\nu$ potential is switched off.  In the QKE solver, where the density matrix is momentum averaged, this is a weaker restriction than it appears: with a single $\rho$, the self-term reduces to $[\rho-\bar\rho,\rho]=-[\bar\rho,\rho]$, so the same-species coupling cancels identically and only the neutrino-antineutrino commutator survives.  Consistently with this, enabling the full system reproduces the adiabatic result at the asymmetry benchmark of Ref.~\cite{Froustey:2021azz}.  Restoring the collective same-species coupling therefore requires momentum resolution, and a momentum-resolved treatment is left for future work.

The QKE solver is integrated on to $0.8\mev$, where the integrated flavor asymmetries stop evolving, in a single restricted role: it carries those asymmetries rotated into the propagation basis, and nothing else.  The rotation uses $U_{\rm C}$, the matrix that diagonalizes the momentum-averaged Hamiltonian the solver itself evolves, the vacuum mass term together with the CP-even thermal potential of the electrons and muons, taken at the temperatures of the step and at the solver's reference momentum $p\simeq2.6\,T_\nu$.  The trajectory that enters the transport is therefore $q_i(\ln a)=[U_{\rm C}^\dagger Q\,U_{\rm C}]_{ii}$.

While the scale factor lies inside the solved range, each transport step is bounded in $\ln a$ by
\begin{equation}
 \epsilon_{\rm tr}^{\rm eff}=\min\!\left(\epsilon_{\rm tr},
                                         \Delta\ln a_{\rm QKE}\right),
 \label{eq:qke-transport-cadence}
\end{equation}
where $\epsilon_{\rm tr}$ is the requested transport step and $\Delta\ln a_{\rm QKE}$ is the step at which the solver was advanced.  Beyond the last solver point the requested step resumes and the matching ends there, while the sampling that assigns propagation labels runs to the end of the transport.

At the end of every step inside that range, once the HNL daughters from the interval have entered the pool, $q_i$ is evaluated at the current scale factor.  Both calculations contain the same HNL decays, so the transport already carries the asymmetry those decays deposited.  The matching therefore keeps the transport's own total and takes only the differences among the three propagation states from the solver,
\begin{equation}
 q^{\rm target}_i=q_i+c,\qquad
 c=\frac13\left(\sum_j q^{\rm tr}_j-\sum_j q_j\right),
 \label{eq:charge-offset}
\end{equation}
which gives $\sum_i q^{\rm target}_i=\sum_i q^{\rm tr}_i$ and leaves every difference $q^{\rm target}_i-q^{\rm target}_j$ equal to the solver's.  The asymmetry the decays deposited stays where the transport put it.

The pool is brought to that target by relabeling, at fixed particle energies and at fixed numbers of neutrinos and of antineutrinos separately.  Moves are made in pairs, $\nu_i\to\nu_j$ together with $\bar\nu_j\to\bar\nu_i$ inside one energy cell, which leaves the CP-even census of that cell unchanged.  Unpaired moves would change it and are switched off.  Each particle moves at most once per step, and the residual that restriction leaves is recorded.

\emph{Photon-heating normalization.}  The mesons among the HNL decay products, the pions and the kaons, are absorbed on nucleons in competition with their own decay.  They therefore couple the evolution of the neutrino and electromagnetic populations to the evolution of the nucleons, and in particular to the baryon-to-photon ratio $\eta_B$~\cite{Akita:2024nam,Akita:2024ork}.  The calculation must track $\eta_B$ through the whole epoch when the HNLs are present.  Its final value is fixed by the CMB, $\eta_B(T_{\rm CMB})=\eta_B^{\rm Planck}$.  Anchoring the ratio at the end leaves it undetermined at earlier times, because relating the two requires the history of the scale factor, and that history depends on the HNL evolution and on the neutrino and electromagnetic populations that the run is computing.  

Define $\zeta_\gamma=(a_fT_{\gamma f})^3/[2.75(a_0T_{\gamma0})^3]$, so $\zeta_\gamma=1$ is standard $e^+e^-$ heating.  The HNL decays deposit electromagnetic energy and heat the photons beyond the standard amount, which dilutes the baryon-to-photon ratio, so the calculation must start from a larger baryon-to-photon ratio for the final value to match the CMB determination.  The transport needs this number before it starts, because it fixes the baryon density it carries through the run, and the run then realizes a photon heating of its own.  Writing $\zeta_{\rm in}$ for the value the transport starts from and $\zeta^{\rm real}(\zeta_{\rm in})$ for the value the run realizes, the normalization at each asymmetry point is the value that reproduces itself,
\begin{equation}
 \zeta^{\rm real}(\zeta_{\rm in})=\zeta_{\rm in}.
 \label{eq:heating-fixed-point}
\end{equation}
The fixed point is solved on an integrated transport that is momentum-averaged, omits the non-thermal tail and the oscillation stage, and uses Born weak rates.  Two passes of it suffice.  The first pass takes $\zeta_{\rm in}=1$ and measures what the decays realize. The departure from unity comes from the electromagnetic energy of the decays, and it is large: $\zeta^{\rm real}=3.30$ at $(m_N,\tau_N)=(0.60\gev,0.069\s)$.  It depends on the asymmetries as well as on $m_N$ and $\tau_N$.  The second pass re-measures it with the meson fates corrected. The remaining feedback runs from $\zeta_\gamma$ to the baryon density, from there to the fraction of mesons that are absorbed on nucleons before they decay, and from there to the division of the released energy between the electromagnetic and the neutrino sector.  Its size is the change between the two passes, $0.2\%$ at that point, so one update closes the integrated transport on itself.  

The production transport is momentum-resolved and realizes a heating of its own, so the normalization the integrated transport returns is not reproduced exactly.  That residual does not move the baryon density at which the abundances are computed.  The nuclear stage sets that density from the photon heating each run realizes, so it works at the CMB baryon-to-photon ratio whatever value the integrated transport returned, and the HNLs have decayed long before the network starts at $86\,$keV, so the baryon-to-photon ratio follows its standard trajectory across nucleosynthesis itself.  The residual moves the baryon density the transport carries, which fixes the fraction of injected mesons absorbed on nucleons before they decay, so it reaches the abundances through the neutron fraction and the injected spectra.  Absorption on nucleons is a small loss channel for the mesons compared with their decay, so the meson population, and with it the conversion rates, barely depend on that density.  Below $1\mev$, where the conversion fixes the neutron fraction, the residual moves the rates by well under a percent.  The condition is not fitted to data: it does not use $\yp$, D/H, or $\neff$.

\subsection{The BBN code}

The nuclear stage is adapted from the BBN framework of Ref.~\cite{EscuderoAbenza:2025tsi}.  Over the temperature range relevant here it already carries the modification of the weak $\pn$ rates by a nonstandard neutrino population, the meson-driven $\pn$ rates, and the contribution of a new relic to the energy density of the Universe.  We extend it in two respects: neutrinos and antineutrinos are carried as distinct populations, and each distribution is carried as a table of arbitrary shape.

The nuclear stage evolves $n,p,{\rm D},{\rm T},{}^3{\rm He},{}^4{\rm He}, {}^7{\rm Li},{}^7{\rm Be}$ through the 12 dominant reactions describing nucleosynthesis up to ${}^7$Be, integrates the weak rates over the transported spectra with the outer-radiative and Coulomb kernels of Ref.~\cite{Pitrou:2018cgg}, and uses modern rates of deuterium burning.  The classification uses only $\yp$ and D/H; we make no precision claim for ${}^3$He or the mass-seven abundances.  Finite-nucleon-mass/weak-magnetism and thermal radiative pieces are not included; a full network or broad nuclear-rate scan would be needed for precision likelihood contours.  The fixed active-neutrino collision coefficients of the production stage remain an approximation; the plasma thermodynamics itself is asymmetry dependent through the five-potential charge redistribution (SuM~\ref{app:production}).

\subsection{Example of the evolution}

\begin{figure*}[t!]
    \centering
    \includegraphics[width=0.49\linewidth]{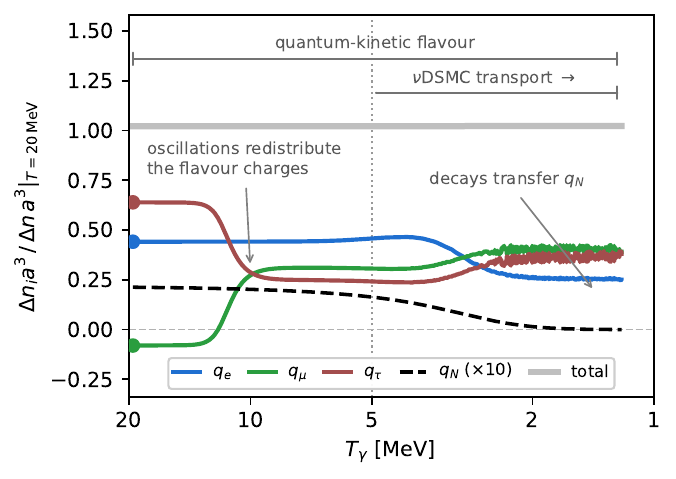}\hfill
    \includegraphics[width=0.49\linewidth]{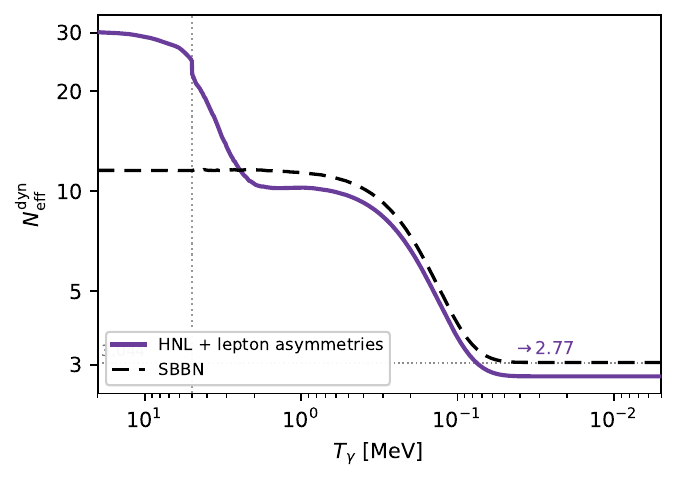}
    \caption{Computed evolution at $(m_N,\tau_N)=(0.60\gev,0.069\s)$.  \emph{Left}: the comoving lepton asymmetries of Eq.~\eqref{eq:comoving-charges}, with the production output at $20\mev$ marked by dots and $q_N$ scaled by ten.  \emph{Right}: the dynamical effective neutrino number of Eq.~\eqref{eq:neff-dynamical}, for the same point (purple solid) and for standard BBN computed with the same chain (black dashed); the dotted horizontal line is $3.044$.  In both panels, the brackets give the temperature range covered by each of the two computational stages and the dotted vertical line marks where the momentum-resolved transport stage begins, at $5\mev$.}
    \label{fig:asym-evolution}
\end{figure*}

To show how the three stages and the BBN code work together, we follow a single point through all of them, $(m_N,\tau_N)=(0.60\gev,0.069\s)$.  Figure~\ref{fig:asym-evolution} traces it from the end of the production stage at $20\mev$ through BBN, in terms of the comoving asymmetries
\begin{equation}
 q_i(T)\equiv\frac{\Delta n_i\,a^3} {\sum_\alpha\Delta n_\alpha a^3\big|_{20\mev}}\,,\qquad \Delta n_\alpha=n_{\nu_\alpha}-n_{\bar\nu_\alpha}\,,\quad \Delta n_N=n_N-n_{\bar N}\,,
 \label{eq:comoving-charges}
\end{equation}
with $i=e,\mu,\tau,N$, and of the dynamical effective neutrino number
\begin{equation}
 \neff^{\rm dyn}(T)\equiv\frac{8}{7} \left(\frac{11}{4}\right)^{4/3} \frac{\rho_\nu(T)}{\rho_\gamma(T)}\,,
 \label{eq:neff-dynamical}
\end{equation}
evaluated on the instantaneous neutrino and photon energy densities, so that it becomes the observable $\neff$ once $e^\pm$ annihilation has completed.  Four features of this evolution are relevant.  The oscillations redistribute the unequal production era flavor asymmetries until they are nearly equal before BBN, the muon and tau components equilibrating quickly while the electron component equilibrates only partially, which is the origin of the residual freedom in how the asymmetry is divided between the electron and tau flavors.  The decays return the HNL asymmetry $q_N$ to the active leptons, where it reappears as the growth of the flavor asymmetries.  Independently of that, the charge-asymmetric meson injection biases the strong $p\leftrightarrow n$ conversion rates.  The total lepton charge is conserved throughout, to $9.6\times10^{-4}$ of itself.  Finally $\neff^{\rm dyn}$, which the primordial degeneracies leave far above its standard counterpart at $20\mev$, is brought back down by the $e^\pm$ annihilation and by the decay heating of the photons: the two curves converge to about a percent of each other below $0.5\mev$, and the surviving difference is $\dneff=-0.280$: at the lowest temperature plotted, $\neff=2.773$ with the HNL source against $3.053$ without it.  The percent-level scatter of the transport is Monte-Carlo noise.  It is a single chain at fixed $L_\alpha$, so it sits at a slightly different member of the family than the scanned solution of Table~\ref{tab:benchmarks-extended}, where $L_\mu$ is re-solved from the helium condition at every step.  This is the cancellation of the main text made explicit.  The flavor asymmetries plotted here come from the QKE solver at every temperature shown, so they run through $5\mev$ without a seam; the change there is that stage~3 begins resolving the spectra in momentum.

\section{Algorithm for locating the regions of Fig.~\ref{fig:money}}
\label{app:scan}

At each grid point $(m_N,\tau_N)$ the asymmetries are located by nested one-dimensional root searches.  The helium condition is solved first: an estimate from the production stage alone initializes $L_\mu$, and a secant iteration of the full calculation then drives $\Delta\yp$ toward zero, stopping once $|\Delta\yp|\leq1.5\times10^{-3}$.  That tolerance exceeds the upper edge of the helium window of Table~\ref{tab:domain-definitions}, so a solution can stop at a positive shift that the window excludes.  The domains are conservative in that direction, and solving to a tighter tolerance can only add points to them.  The tau asymmetry $L_\tau$ is set to several values, chosen so that the residual $\dneff$ covers the windows of the standard $\neff$ and $\neff$ deficit regions (Table~\ref{tab:domain-definitions}).  For each of them, $L_e$ runs over both signs, with $L_\mu$ determined by the helium condition.  At every step of this scan, we evaluate the ratio $\mu_*(T_{\rm CEP})/\mu_{\rm CEP}$ that enters the first-order criterion of Eq.~\eqref{eq:fo-criterion}, the temperatures $T_c(k)$ at which $\mu_*$ crosses the assumed first-order lines, and the condensation margin $\max|\mu_Q|/m_\pi$ (SuM~\ref{app:qcd}).  We also compute the enhancement $R_\nu$ of the relic neutrino number density (SuM~\ref{app:cnub}).  Together, these scans over $L_\tau$ and $L_e$ sample the one-parameter family of asymmetries at each $(m_N,\tau_N)$, whose members differ in the division of the net asymmetry between the electron and tau flavors.  Within the explored asymmetry ranges, the nested searches encountered no secondary solutions: at fixed $(m_N,\tau_N)$, the map from $(\yp,\dneff)$ back to the asymmetries is single-valued up to that division.

\subsection{Statistics of the transport}

Every quoted observable is a difference $\Delta O\equiv O_{\rm HNL+LFA}-O_{\rm SBBN}$ against a standard-BBN calculation in which both the HNL population and the primordial flavor asymmetries are set to zero.  The two are run from the same random numbers, so their Monte-Carlo noise is common and largely cancels in the difference, which is therefore resolved well below the scatter of either one alone.  Candidate solutions are first located in a coarse pass with $10^4$ neutrinos per species.  The asymmetries of every quoted solution are then determined with four independent runs at $3.2\times10^4$ particles per species, where the transport is converged also at strong $\tau$-degeneracy (SuM~\ref{app:convergence}).  Each candidate is finally re-evaluated with twelve further runs at the same number of particles per species, independent of those used to determine it, and is accepted only if it is not excluded at one standard error by any window of Table~\ref{tab:domain-definitions}, that is, if the interval of one standard error about the mean overlaps the window.  This is an overlap criterion, weaker than requiring the whole interval to lie inside.  The residuals of the accepted benchmark solutions are listed in Table~\ref{tab:benchmarks-extended}.

\subsection{How sharply the observables fix the asymmetries}

The scan itself measures how sharply the observables select the asymmetries.  Across the scan, the local responses have median $|\partial\yp/\partial L_\mu|\simeq15$ and $|\partial\dneff/\partial L_\tau|\simeq6$, so the $\yp$ band corresponds to $\delta L_\mu\simeq\pm2\times10^{-4}$ and the $\neff$ band to $\delta L_\tau\simeq\pm8\times10^{-3}$, at the percent level of the selected values, while moving asymmetry between the electron and tau flavors leaves the bands intact over $\Delta L_e\gtrsim0.1$ (SuM~\ref{app:detailed}).  These widths translate the adopted observational windows into local allowed widths in asymmetry space within the fixed calculation.  They do not include any theoretical uncertainty, and the uncertainties introduced by the numerical settings and by the stage matching are quantified separately (Table~\ref{tab:numerical-stresses}).

\subsection{Domain definitions and grid}

The regions constructed below end at $\tau_N\simeq1\s$.  At such lifetimes, the HNLs act on BBN and the CMB only through the $\pn$ conversion, $\neff$, the helium abundance, and the baryon-to-photon ratio (SuM~\ref{app:framework}); the remaining effects, the altered expansion before recombination~\cite{Vincent:2014rja}, the photodisintegration of the nuclei~\cite{Domcke:2020ety}, and the spectral distortions and modified anisotropies of the CMB~\cite{Poulin:2016anj,Diamanti:2013bia}, need $\tau_N$ above about $10\s$, while fewer than a fraction $e^{-200}$ of the population survives to the onset of nucleosynthesis near $200\s$; we therefore do not consider them.

The parameter scan classifies every sampled grid point against the classification windows of Table~\ref{tab:domain-definitions}.  Every condition is applied to $\Delta O$ as defined above, the change relative to the standard-BBN calculation with no HNLs and no primordial asymmetries.  The deuterium window combines the one-percent measurement error with the comparable spread among the theoretical predictions and, as for helium, is taken at two sigma, which gives $|\Delta(\mathrm{D/H})|/(\mathrm{D/H})\leq0.031$.  For helium the theoretical error is small, and the measured value lies below the standard prediction, so we refer the window to the measurement.  Standard BBN gives $\yp\simeq0.2471$, so the shift that reproduces the observed value is $-0.0013$, and we accept shifts within $2\sigma$ of it, $-0.0038\leq\Delta\yp\leq0.0013$.  The interval is asymmetric because the standard prediction sits $1.0\sigma$ above the measurement, which leaves more room below.  The confidence level also sets the floor of the $\neff$ deficit window, through the correlation $\Delta({\rm D/H})/({\rm D/H})\simeq0.11\,\dneff$: the floor is the deuterium window divided by $0.11$, which gives $-0.28$ at two sigma and $-0.14$ at one.  The two windows are therefore the same construction at two confidence levels, and we adopt two sigma so that deuterium and helium are treated alike.  Repeating the classification with the one-sigma deuterium window, and with the floor it implies, leaves the standard $\neff$ region essentially unchanged, shrinks the $\neff$ deficit region by about a quarter of its area, and the GW target region by almost half.  Every region remains populated over the full mass range and keeps the lifetime interval $0.015$ to $0.29\s$ inside the SHiP reach, so neither the mechanism nor the experimental opportunity depends on the choice of confidence level.  The scan covers 9 masses from $0.25$ to $0.95\gev$ and 24 lifetimes from $0.0075$ to $1.1\s$, spaced most densely where the regions lie; the lifetime range extends past their upper edge.  That edge is where a second effect adds to the meson-driven conversion: HNLs decay into high-energy neutrinos, directly or through the decays of the metastable muons and mesons they inject, and at these late times the plasma no longer thermalizes them: after $e^\pm$ annihilation, their scattering rate on the neutrino background has fallen to $10^{-3}$ of the expansion rate, so their spectrum is never driven back to a thermal shape.  These neutrinos reach BBN by converting protons into neutrons.  Each one does so rarely, eight orders of magnitude more rarely than it scatters.  The fraction of protons converted decides whether this matters, and that fraction reaches $\sim10^{-5}$.  Deuterium makes up only about $2.5\times10^{-5}$ of hydrogen, so converting protons at that level changes D/H substantially.  The population is never replenished and never thermalized, so it simply redshifts while oscillations keep converting it into the electron flavor.  It therefore goes on converting protons into neutrons after nucleosynthesis has begun.  For example, for $m_N=0.3\gev$ and $\tau_N=0.5\s$ with no asymmetry, $3\%$ of the $\bar\nu_e$ still lie above the $\bar\nu_e\,p\to n\,e^+$ threshold at $T=80\,$keV, just after the nuclear network starts at $86\,$keV, against $2.5\times10^{-6}$ for a thermal spectrum of the same mean energy.

\emph{The deuterium prediction.} The $\neff$ deficit region of Fig.~\ref{fig:money} depends on the adopted theoretical prediction for D/H: of the two in the literature, we adopt the higher~\cite{Pisanti:2020efz,Gariazzo:2021iiu}, which agrees with the measured abundance. With the lower~\cite{Pitrou:2018cgg,Pitrou:2020etk}, the correlation $\Delta(\text{D/H})/(\text{D/H})\simeq0.11\,\dneff$ selects positive $\dneff$ instead: the deficit window closes, an excess window opens, and the solutions with positive $\dneff$ at the short lifetimes of the domains populate it. The standard $\neff$ region keeps the same agreement with D/H as SBBN itself, and the relaxation of the meson-driven bound and the solved asymmetries do not change.

$\yp$ is fixed once the deuterium bottleneck breaks and the surviving neutrons are locked into $^4$He, so a neutron created after that stops at deuterium and the late conversions add deuterium while leaving helium alone.  Removing the non-thermal part of the spectrum at fixed neutrino energy density confirms it: at the same point, the D/H excess falls by a factor of ten while $\neff$ moves by one percent.  The late conversions add deuterium, and the inverse process finds no neutrons left to act on. Helium is fixed before they occur, so the muon flavor component that the helium window sets does not act on them. Across the asymmetries sampled here, the excess grows as a steep power of the lifetime and leaves the deuterium window at the upper edge of the domains, and the division of the net asymmetry between the electron and tau flavors was scanned on a coarse grid.  We verified the attribution by disabling all $p\leftrightarrow n$ interconversion except free neutron decay after the network switch-on at fixed transport history: the D/H excess collapses to the standard value while $\yp$ is unaffected.  The two ends of the domain terminate for different reasons.  At the light end deuterium stops it, and the stop is physical: at $0.3\gev$ and $\tau_N=0.8\s$ the $\neff$ condition is still met, $\dneff=+0.03$, while D/H exceeds its window by a factor of three, and across the asymmetries sampled at that point the excess varies by less than a fifth, so no member of the family recovers it.  At the heavy end the $\neff$ condition binds first, and how far it can be pushed is set by the convergence of the momentum-averaged solver, a numerical limit; the boundaries quoted above $0.7\gev$ are therefore conservative.  Each grid point is classified independently: no interpolation or extrapolation enters the classification, and a point whose asymmetries were not located enters no domain.  The boundaries drawn in Fig.~\ref{fig:money} are built from those classifications.  For each domain, the boundary is drawn from the longest and the shortest lifetime that meets its requirements at each mass.  A heavier HNL with the same lifetime injects more energy, so the upper end of each domain cannot move to longer lifetimes as the mass increases.  The boundary at a given mass therefore extends to the longest lifetime classified in the domain at that or any larger mass.  The sampled lifetimes differ by about $19\%$, so the edges are finally smoothed across neighboring masses.
\section{Correctness and convergence of the calculation}
\label{app:validation}

The checks below test the three stages of SuM~\ref{app:framework} and the BBN code that follows them.  They fall into two groups.  The first asks whether the calculation is correct, comparing it against analytic limits, thermodynamic identities, conservation laws, and independent calculations.  The second asks whether the numerical settings are converged, varying the step size, the particle count, and the stage-matching temperature.

\subsection{Cross-checks}

The checks passed by the calculation are collected here, stage by stage.

\begin{itemize}[leftmargin=1.2em,itemsep=2pt,topsep=3pt,parsep=0pt]
\item \emph{Production and thermodynamics.}  The framework reproduces the standard analytic results and thermodynamic identities of an ideal gas.  The relation between the neutrino-antineutrino number asymmetry and the dimensionless chemical potential, $\Delta n=T^3(\xi+\xi^3/\pi^2)/6$ for a massless fermion, is recovered at the $10^{-15}$ level.  The Gibbs relations, which give the number and entropy densities as derivatives of the pressure, $n=\partial P/\partial\mu$ and $s=\partial P/\partial T$, hold at the $10^{-8}$ level for fermions and bosons alike and in both the degenerate and the dilute regime.  The total lepton charge is conserved to machine precision.  The charge-symmetric limit reproduces, at the $2\times10^{-4}$ level in the yield, a calculation in which each flavor charge is stored in its own neutrino species alone, and the charge-neutrality trajectories at the QCD transition epoch agree with an independent free-quark-gas calculation once the entropy-normalization conventions are matched~\cite{DiClemente:2025awt}.
\item \emph{Asymmetry redistribution at large flavor asymmetries.}  With no HNLs present and the asymmetries of interest, our five-potential solve reproduces the independent calculation of Ref.~\cite{Akita:2025txo} to $0.2$ to $3\%$ in the degeneracies $\xi_\alpha$ below the QCD crossover, and to $2$ to $3\%$ at the production temperatures, once the entropy-normalization conventions and the modeled particle content are matched.  Both comparisons cover the regimes that matter: the first sets the asymmetries passed to the oscillation stage, the second sets the particle-antiparticle ratio of the HNL population.  The two treatments differ in content: we include the bottom quark and that reference does not, which changes $\xi_\mu$ by $39\%$ at $T\gtrsim2\gev$.  The bottom quark is relativistic there and carries about an eighth of the entropy, so including it changes the number density that corresponds to a given $L_\alpha$.  Of the three degeneracies, $\xi_\mu$ is the smallest, the muon flavor charge being carried mostly by the charged muons, so the same shift in normalization moves it by tens of percent while $\xi_e$ and $\xi_\tau$ move by about five.
\item \emph{Sensitivity to the QCD equation of state.}  Replacing the ideal-quark description by a model calibrated to reproduce the chemical potentials of Ref.~\cite{Akita:2025txo} across $0.5$ to $2\gev$ changes the particle-antiparticle ratio of the produced population by $7\%$ at $(0.55\gev,0.03\s)$, a representative short-lifetime point, at which the HNLs therefore decouple late.  The helium condition is inverted against that ratio, so a systematic shift in it is absorbed by re-solving $L_\mu$ and moves the selected asymmetries and leaves the abundances in place; the full node solve has not been repeated under the alternative description.
\item \emph{Thermodynamics at the QCD transition epoch.}  The charge redistribution across the crossover uses two ideal-gas descriptions, free quarks and gluons above $250\mev$ and a truncated hadron-resonance gas below $160\mev$, with exact Fermi-Dirac and Bose-Einstein integrals on both sides.  Only the hadron-resonance gas enters the QCD criterion of the GW target region, pion condensation with $|\mu_Q|\geq m_\pi$ at $T\leq140\mev$ (SuM~\ref{app:qcd}).  We have also compared the hadron-resonance-gas description against a lattice-susceptibility description of the same region, in which the QCD pressure is expanded to second order in the chemical potentials, and find the charge chemical potential to agree to about ten percent at the asymmetries of the solutions and at the temperatures where the onset of pion condensation is evaluated.  Along the family of asymmetries at $(0.6\gev,0.14\s)$, a change of $\mu_Q$ by that amount shifts the value of $L_e$ at which condensation sets in by about $0.03$.  At these temperatures $\mu_Q/T$ approaches unity, so neither description is reliable there.  This uncertainty affects the extent of the GW target region, whose amplitude requires calculations of bubble nucleation and wall dynamics that we do not perform.  It does not reach $\yp$, D/H, or $\neff$: below the crossover, the flavor asymmetries are carried by neutrinos, $e^\pm$, and photons alone, so the conserved $L_\alpha$ fix the degeneracies at $20\mev$ whatever description holds at the crossover, and the sensitivity of the produced population to the equation of state is quantified in the previous item.
\item \emph{Division of the injected energy among the flavors.}  Stage~2 carries a single neutrino temperature, so the way the injected energy is distributed between flavors is not resolved there.  We bound the effect by re-initializing stage~3 with the opposite extreme, in which each species keeps the energy that its own decay channels deposit in it.  At $(0.55\gev,0.03\s)$, this shifts $\dneff$ by $-0.017\pm0.011$, $\yp$ by $+0.0003\pm0.0005$, and D/H by $(-0.12\pm0.16)\%$.  We take the difference between the two prescriptions as the size of the uncertainty that the matching between the stages introduces.
\item \emph{Stage-2 coupling.}  Without the HNL source the calculation returns $T_\nu=T_\gamma$ at $5\mev$ within $10^{-3}$, $X_n=0.4358$ against the weak-equilibrium value $0.436$, the standard elapsed time of $0.0277\s$ between $20$ and $5\mev$, and per-step conservation of lepton number at the $10^{-13}$ level.  With the HNL source included, each injection preserves the six number increments and the total injected neutrino energy to the same precision.  The temperature at which stage~3 begins is not a physical scale, so the answer must not depend on it: moving it from $5$ to $5.5\mev$ shifts $\yp$, D/H, and $\neff$ by amounts that are not resolved from zero (Table~\ref{tab:numerical-stresses}).
\item \emph{End-to-end SBBN.}  Without the HNL source the main abundances agree with precision standard BBN~\cite{Pitrou:2018cgg,Pitrou:2020etk,Pisanti:2020efz} at about $0.5\%$ in $\yp$ and below $1\%$ in D/H.  Since every quoted observable is the difference between the calculation with and without the HNLs, a constant offset of that baseline cancels and does not enter the result.
\item \emph{HNLs in the scenario with zero lepton flavor asymmetries.}  Here the population is charge symmetric, $r=1.000$, and the calculation reproduces the results of Refs.~\cite{Boyarsky:2020dzc,Sabti:2020yrt} in two regimes: masses above the pion production threshold, where the BBN constraints are typically dominated by the meson-driven conversion, and below it, where the main effect is the impact on the neutrinos.\par At $m_N=0.3\gev$, we recover the bound $\tau_N<0.02\s$ of Ref.~\cite{Boyarsky:2020dzc} under two assumptions: the window on $\yp$ used there and no neutrino spectral distortions.  Including them strengthens it to $\tau_N\simeq0.013\s$.  The HNLs with $\tau_N\simeq0.013$ to $0.02\s$ are therefore constrained by the distortions, which accumulate during the decays at $T\gtrsim3\mev$ and survive at lower temperatures, where neutrino equilibration is already slow.  At longer lifetimes the conversion dominates: switching it off leaves the full helium effect at $\tau_N=0.01\s$, where the injected mesons are still too few to matter, a seventh of it at $0.02\s$, and a tenth at $0.05\s$, consistent with Ref.~\cite{Sabti:2020yrt}, which reports that including this channel would strengthen their bounds by about a factor of two.\par In the second regime, we compare $\neff$ against Ref.~\cite{Sabti:2020yrt} at the mass and mixing pattern of its Fig.~5, $m_N=0.2\gev$ coupled to the electron flavor, and find very good agreement across $\tau_N=0.01$ to $0.05\s$.  The treatments still differ in how the metastable species that the decays inject, $\mu^\pm$, $\pi^\pm$, and $K$, are evolved: that reference lets them decay promptly, up to the kinetic energy the charged species lose, while we also follow their annihilation on one another and their interactions with nucleons.  That difference changes how the released energy is divided between the neutrino and the electromagnetic sector, and at this mass it does not move $\neff$ beyond that agreement.

\item \emph{Detailed balance in the production stage.}  The production solver relaxes the HNL occupation at each momentum towards the active occupation at that momentum, so raising the mixing at fixed lifetime must drive the abundance to its equilibrium value.  At $m_N=0.3\gev$, with the decay switched off so that production is isolated, the comoving abundance saturates at $Y_N=Y_{\bar N}=7.29\times10^{-3}$ and is independent of $|U|^2$ between $10^{-5}$ and $0.3$, the population having decoupled at $T=263\mev$.  The equilibrium value there, $Y=g\,n_{\rm FD}(m_N,T)/s$ evaluated with the exact massive Fermi-Dirac integral, $g=2$ spin states per species and $g_{*s}=46.2$, is $7.23\times10^{-3}$; the solver reproduces it to $0.85\%$.  The same test fixes the degeneracy: a Dirac HNL carries two spin states for $N$ and two for $\bar N$, and the abundance agrees only with that counting.

\item \emph{Entropy release, against an independent calculation.}  The decays heat the photon bath and dilute the baryon-to-photon ratio by $\zeta\equiv[(aT)_{\rm SBBN}/(aT)_{\rm HNL}]^3$.  At $(m_N,\tau_N)=(0.5\gev,0.04\s)$ with no asymmetries our chain gives $\zeta=0.406$, against $\zeta\simeq0.41$ in Fig.~5 of Ref.~\cite{Boyarsky:2020dzc}, an agreement of $1\%$.  The ratio is converged, moving by $0.3\%$ over the last half of the trajectory.  In the same run the comoving HNL number follows $e^{-t/\tau_N}$ to $1\%$ over more than one lifetime, with the fitted lifetime within $1.4\%$ of the input.  This tests the abundance, the decay law, and the electromagnetic energy release together, and unlike a comparison of $\neff$ it is insensitive to the treatment of the metastable species, which the two calculations handle differently.

\item \emph{The rule that carries the QKE asymmetries into the transport.}  That rule was tested by replacing it with an alternative.  Under the first, the particles carry flavor labels while the QKE trajectory is active, and one operation at the end converts the population into propagation states with probabilities $|U^{\rm m}_{\alpha i}(E,T)|^2$.  Under the second, the particles carry propagation states throughout and the solver contributes only $q_i=[U_{\rm C}^\dagger Q\,U_{\rm C}]_{ii}$, matched at every step.  At $(m_N,\tau_N)=(0.775\gev,0.5\s)$ with $L_e=0.357$ and $L_\tau=0.271$, fifteen seeds per rule at $3.2\times10^4$ particles per species, the two are compared seed by seed.  The abundances are unchanged within the Monte-Carlo scatter, so the $\yp$ and D/H criteria of Table~\ref{tab:domain-definitions} are insensitive to the choice.  The neutrino observables move a little: $\Delta\neff=+0.017\pm0.004$, a third of the $|\dneff|=0.05$ window of the standard region.  The domains are not evaluated at fixed asymmetries: $L_\mu$ is re-solved from the helium condition everywhere, so the shift is absorbed along the family, $\delta L_e\simeq0.006$, well inside one step of the $L_e$ grid, and a drawn edge can retreat only by that margin.  Applying a rigid $+0.021$, larger than the measured value, across the scan and repeating the classification confirms it: all four regions and their SHiP overlaps survive; the largest change, in the GW target, is a loss of a fifth of the points and a seventh of those within reach.  The pairing correlation runs from zero for D/H to $0.76$ for $\neff$, which is why the deuterium error stays wide while the $\neff$ error tightens.  The results throughout use the second rule, in which flavor enters through the collision amplitude and no draw is made before the collision is accepted.

\item \emph{The charge the transport receives.}  The QKE charge matrix is diagonal in the propagation basis to better than $4\times10^{-3}$ over the transported range, so $[U_{\rm C}^\dagger Q\,U_{\rm C}]_{ii}$ returns the propagation-state asymmetries themselves.  Read back from the transported labels, the electron asymmetry is reproduced to $0.07\%$.  The residual sits in the muon-tau split, redistributed by about $30\%$ with their sum conserved to $0.1\%$, and it follows from evaluating the propagation basis at the single reference momentum of the momentum-averaged solver, in the sector where those two states are close to degenerate.  The electron asymmetry sets the $\pn$ rates.

\item \emph{The transport in isolation.}  In a comoving volume with no asymmetry and no source, the flavor populations relax toward equality, the spread falling from $30\%$ to $2\%$ of the mean while the net asymmetry is conserved exactly.  Starting from equal populations, no asymmetry appears: over eight seeds the mean asymmetry per species sits within half a standard error of zero.

\item \emph{The standard limit and the step size.}  With no HNL the calculation returns $\neff=3.058\pm0.009$ and $\yp=0.2467\pm0.0007$, against $3.044$ and $0.2470$.  Halving the transport step changes the helium shift by $-3\times10^{-5}\pm7\times10^{-4}$, inside the $+0.0013$ margin of the asymmetric window.

\end{itemize}

\subsection{Convergence tests}
\label{app:convergence}

A convergence test asks whether a result depends on a setting that carries no physical meaning.  Such settings are chosen for numerical convenience alone, so any dependence of $\yp$, D/H, or $\neff$ on them would be an artifact.  We vary each in turn and require the induced change to be small compared with the acceptance band that defines the domains.

\emph{Step size and particle number.}  At $(0.5\gev,0.05\s)$, $(0.7\gev,0.022\s)$, and $(0.6\gev,0.14\s)$, we compare the reference setting against a halved quantum-kinetic step and against $64000$ particles per species, twelve paired realizations per comparison at the same random numbers, with the asymmetries held at the benchmark solution so that each difference isolates the numerical setting.  Table~\ref{tab:numerical-stresses} gives the resulting changes in the observables, together with the response to moving the start of stage~3 from $5$ to $5.5\mev$ at two of the three points.  The engine refuses a handoff above $6\mev$, where the two equations of state it matches stop agreeing, so $5.5\mev$ varies the same setting while staying on the valid side of that boundary.  Only the helium shift at $(0.60\gev,0.143\s)$ under doubled statistics is resolved from zero, at $2.4\sigma$; every other entry is consistent with zero and bounds the corresponding numerical error from above.  Taking $|\text{mean}|+\text{s.e.m.}$ as that bound and comparing it with the half-width of the matching window of Table~\ref{tab:domain-definitions}, the largest reaches $44\%$ of it, the helium bound at that same point.

\emph{Transport statistics at strong degeneracy.}  The particle number matters most where a single flavor is strongly degenerate.  At $(0.55\gev,0.17\s)$, where the selected neutrino distribution reaches $\xi_\tau\simeq3.2$, raising the particle number at the same random numbers gives $\dneff=+0.09$, $+0.23$, and $+0.24$ at $10^4$, $3.2\times10^4$, and $6.4\times10^4$ particles per species.  The cause is the sampled occupancy that enters the Pauli-blocking factors, which carries $\mathcal{O}(1/\sqrt{N})$ noise: in a nearly filled Fermi sea, the blocking factors are clipped at zero, and the clipping systematically lowers $\neff$ at small $N$.  The effect is negligible for $|\xi|\lesssim1$ and converged at $3.2\times10^4$ particles per species, with a residual $\lesssim0.02$.  That setting therefore defines every quoted number, and the $10^4$ pass is used only to locate candidate solutions.  In the sequential transport used here, every accepted collision updates the binned occupancies before the next collision is proposed, and a rejected proposal restores their pre-collision values.  The transport estimators are detailed in Ref.~\cite{Akita:2026xwu}.

\emph{Instantaneous flavor averaging.}  Instantaneous averaging cannot replace the flavor evolution.  With no HNL source, the compensated pattern $(\xi_e,\xi_\mu,\xi_\tau)=(0.78,-0.78,0)$ [Eq.~\eqref{eq:diagnostics}] nearly equilibrates, whereas $(3,-3,0)$ stays substantially flavor asymmetric and shortens the elapsed $20\to5\mev$ time to $0.01819\s$.  Instantaneous averaging maps both patterns onto the same equilibrated state and erases that difference.  Degeneracies of $|\xi|\simeq6$ lie outside the range over which the treatment has been tested.

\begin{table}[t!]
\caption{Response of the observables when a numerical setting is varied, at the same random numbers and the same asymmetry point, relative to the reference setting: $32000$ particles for each of the six neutrino and antineutrino species at the start of stage~3, $d\ln a=0.01$, and stage~3 starting at $5\mev$.  Each point is evaluated at its own asymmetries held fixed, so the entries isolate the numerical setting.  Twelve paired realizations per entry.  D/H shifts are percentage points.  Errors are standard errors of the mean across the paired realizations.  Only one entry is resolved from zero at all, the helium shift at $(0.60,0.143)$ under doubled statistics at $2.4\sigma$, which is unremarkable among eight entries; every other entry is consistent with zero and bounds the corresponding error from above.}
\label{tab:numerical-stresses}
\begin{ruledtabular}
\begin{tabular}{ccccc}
$(m_N\,[\mathrm{GeV}],\tau_N\,[\mathrm{s}])$ & variation & $\delta\yp$ & $\delta[\Delta({\rm D/H})/({\rm D/H})]$ [pp] & $\delta\neff$ \\
\hline
$(0.50,0.0484)$ & halved quantum-kinetic step & $-0.00011\pm0.00030$ & $-0.026\pm0.097$ & $-0.0002\pm0.0051$ \\
$(0.50,0.0484)$ & double transport statistics & $-0.00001\pm0.00026$ & $+0.022\pm0.079$ & $+0.0023\pm0.0047$ \\
$(0.60,0.143)$ & halved quantum-kinetic step & $+0.00062\pm0.00045$ & $+0.232\pm0.126$ & $+0.0087\pm0.0081$ \\
$(0.60,0.143)$ & double transport statistics & $+0.00080\pm0.00033$ & $+0.170\pm0.104$ & $-0.0020\pm0.0071$ \\
$(0.70,0.022)$ & halved quantum-kinetic step & $+0.00063\pm0.00035$ & $+0.085\pm0.107$ & $-0.0053\pm0.0058$ \\
$(0.70,0.022)$ & double transport statistics & $+0.00003\pm0.00033$ & $-0.006\pm0.085$ & $-0.0011\pm0.0044$ \\
$(0.50,0.0484)$ & stage~3 starts at $5.5\mev$ & $-0.00059\pm0.00042$ & $-0.163\pm0.122$ & $-0.0033\pm0.0062$ \\
$(0.60,0.143)$ & stage~3 starts at $5.5\mev$ & $+0.00037\pm0.00042$ & $+0.105\pm0.137$ & $+0.0018\pm0.0089$ \\
\end{tabular}
\end{ruledtabular}
\end{table}

\section{The cosmic neutrino background observable}
\label{app:cnub}

$\neff$ measures the energy density of the neutrinos, so a realization that leaves it close to its standard value can still carry a different neutrino number density.  The spectral distortions from the decays and the flavor asymmetries move the mean neutrino energy and the number density in opposite directions, and the product of the two, to which $\neff$ responds, can barely change while each factor does.  We characterize the cosmic neutrino background by the enhancement of the relic neutrino number density,
\begin{equation}
 R_\nu=\frac{\left[\sum_{\alpha=e,\mu,\tau} n_{\nu_\alpha}/n_\gamma\right]_{\rm HNL}} {\left[\sum_{\alpha=e,\mu,\tau} n_{\nu_\alpha}/n_\gamma\right]_{\rm SBBN}},
 \label{eq:cnub-number-ratio}
\end{equation}
computed from the neutrino spectra at the end of the transport stage.  Neutrino capture on a nucleus, $\nu_e+(A,Z)\to e^-+(A,Z+1)$, relevant for experiments such as PTOLEMY~\cite{Betti:2019ouf}, acts on neutrinos alone, which is why $R_\nu$ counts only them.  In the asymmetric $\nu$ sea the two components differ strongly, so this distinction matters here in a way it would not for a symmetric relic $\nu$ sea.

The capture rate on a non-relativistic relic is governed by $\sigma v$, the capture cross section times the velocity.  For a non-relativistic neutrino, this product is nearly independent of the relic kinetic energy, so the leading homogeneous rate follows the number density~\cite{Betti:2019ouf}.  Equation~\eqref{eq:cnub-number-ratio} is a ratio of homogeneous cosmological densities, and two further inputs turn it into a local capture rate.  The first is the sensitivity of a $\nu_e$ detector to the flavor content of the sea, which does not weigh the three flavors equally.  The capture rate on tritium is proportional to the weighted density $\sum_i|U_{ei}|^2n_{\nu_i}$, in which the number density of each mass eigenstate $\nu_i$ is multiplied by its electron flavor fraction $|U_{ei}|^2$; for Dirac neutrinos, the right-helicity neutrino states are essentially unpopulated in the relic sea, so the rate counts the left-helicity neutrinos~\cite{Long:2014zva}.  The transport stage ends with the neutrinos in mass eigenstates, so the computed spectra give this weighted density directly.  It rises less than $R_\nu$: most of the surviving asymmetry is in the muon and tau flavors, which the capture rate counts only through the electron fractions $|U_{ei}|^2$, and the excess persists even where the $\neff$ deficit vanishes.  For example, at $(0.6\gev,0.14\s)$ it exceeds its standard value by about $40\%$ at $R_\nu=1.64$, and at $(0.5\gev,0.05\s)$, where $\dneff\simeq0$, by about $35\%$ at $R_\nu=1.45$.  The division of the surviving asymmetry between the muon and tau flavors is the weakest input here: the two stages of the calculation redistribute it by about $30\%$ (SuM~\ref{app:validation}), so the weighted density is less certain than $R_\nu$.

In standard cosmology, $100\,$g of tritium captures about four relic neutrinos per year if they are Dirac particles~\cite{Long:2014zva}.  Detecting them requires an energy resolution below the neutrino mass and an amount of tritium beyond current technology~\cite{Betti:2019ouf}.  The zero-point motion of tritium atoms bound to graphene spreads the energies of the emitted electrons over more than that resolution~\cite{Cheipesh:2021fmg,PTOLEMY:2022ldz}.  Observing an enhanced capture rate is therefore a long-term target for experiments such as PTOLEMY.  Without HNLs, a rise of about $30\%$ in that rate requires a neutrino degeneracy that increases $\neff$ by about $0.1$~\cite{Long:2014zva}.  In our solutions, the HNL decays offset this shift, and $\neff$ can stay standard, as at the first benchmark.

The second input is the Galactic clustering of the relic $\nu$ sea, which for the masses allowed by cosmology enhances the local density by a few tens of percent~\cite{Ringwald:2004np}, comparable to the enhancement studied here.  That factor is predicted once $m_\nu$ is known~\cite{Ringwald:2004np}, so it is not a free normalization that could be traded against the enhancement.  It depends on the momentum distribution, but the $\nu$ sea here remains a Fermi-Dirac distribution carrying a degeneracy, with a subdominant non-thermal tail on top of it.  A degeneracy raises the number density much faster than it changes the shape: across the degeneracies of Table~\ref{tab:benchmarks}, the number density moves by factors of several while the mean momentum, in units of the neutrino temperature, rises by at most about a fifth.  The clustering factor is therefore close to its standard value, and converting Eq.~\eqref{eq:cnub-number-ratio} into a local rate rescales the enhancement modestly.  Relic energy affects the capture-electron line position and resolvability more directly than the leading integrated rate.  Direct capture and structure formation are therefore complementary to the energy-density measurement of $\neff$.

\section{Model realization example}
\label{app:model}

Here, we ask what the scenario of the main text requires of a particle-physics setup, and exhibit one construction that meets those requirements.  Three conditions must hold:
\begin{itemize}
\item[--] an HNL charge, the excess of $N$ over $\bar N$, that survives from the production stage until the decays;
\item[--] asymmetries of the active flavors, $L_\alpha\equiv(n_{\nu_\alpha}-n_{\bar\nu_\alpha}+n_{\alpha^-}-n_{\alpha^+})/s$, of order $0.1$, with protection from leakage into baryons during the sphaleron epoch at $T\simeq100\gev$;
\item[--] the generation completed before mixing brings the HNLs into equilibrium at $T\simeq10\gev$.
\end{itemize}
These are the boundary conditions of the kinetic calculation of SuM~\ref{app:production}, which builds the population and its asymmetry from the active-lepton plasma exactly as computed there.

Let us elaborate on the first of them.  The scenario needs the HNL asymmetry $L_N$ to survive until the HNLs decay, since $N$ and $\bar N$ inject pions of opposite charge.  A Majorana mass splitting $\delta m$ makes $N$ and $\bar N$ oscillate into each other, and at least $90\%$ of $L_N$ remains if
\begin{equation}
 \delta m\lesssim0.3\max(\hbar/\tau_N,V_N),
 \label{eq:dirac-condition}
\end{equation}
with $V_N$ the matter potential of the plasma discussed below, taken at $t=\tau_N$; across the scenarios of Fig.~\ref{fig:money}, $\delta m\lesssim10^{-14}$ to $10^{-12}\,$eV follows.  A $\delta m$ this small calls for a symmetry: a lepton number that counts $N$ positively and $\bar N$ negatively, conserved at that accuracy, keeps the pair Dirac and the charge intact.

How demanding this is depends on where the active neutrino masses come from.  If the HNL pair generated them, as in the minimal seesaw, the Higgs mechanism would split it by an amount of order $|m_3-m_2|\simeq0.04\,$eV, and the condition would require a cancellation between two unrelated contributions to one part in $10^{9}$ to $10^{12}$~\cite{Drewes:2019byd}.  Here, the pair carries an exactly conserved lepton number and plays no part in the active neutrino masses, so $\delta m=0$, stable under radiative corrections.  Low-scale seesaw models invoke an approximate version of this symmetry to reach large mixings~\cite{Shaposhnikov:2006nn,Kersten:2007vk}; the dark matter and baryogenesis solutions of the $\nu$MSM need a breaking of about $10^{-11}$ of $m_N$~\cite{Canetti:2012kh}.  Violation of lepton number elsewhere leaks into the pair: a Majorana mass $m_{\mu\mu}$ of the active neutrinos, generated above $m_N$, induces $\delta m=|U_\mu|^2|m_{\mu\mu}|$.  Since $V_N$ also scales with $|U_\mu|^2$, the requirement becomes $|m_{\mu\mu}|\lesssim0.3\,V_{\nu_\mu}$ at the decay epoch, about $10^{-4}\,$eV.  For normal ordering and $\Sigma m_\nu<0.12\,$eV, the oscillation data give $|m_{\mu\mu}|\geq0.01\,$eV~\cite{Esteban:2024eli}, and in the inverted case $m_{\mu\mu}$ vanishes only with tuned phases.  The active neutrinos are therefore Dirac, with masses from the Yukawa couplings of further right-handed states.  A Weinberg operator suppressed by the Planck scale gives $|m_{\mu\mu}|\lesssim3\times10^{-5}\,$eV and is harmless.

The potential itself does not spoil the Dirac nature: forward scattering on the asymmetric plasma shifts the energies of $N$ and $\bar N$ in opposite directions by $V_N\simeq|U_\mu|^2V_{\nu_\mu}$, with $V_{\nu_\mu}=\sqrt2G_F(2\Delta n_{\nu_\mu}+\Delta n_{\nu_e}+\Delta n_{\nu_\tau})$ and $\Delta n_{\nu_\alpha}=n_{\nu_\alpha}-n_{\bar\nu_\alpha}$.  Being diagonal in lepton number, it drives no $N\leftrightarrow\bar N$ transitions; it suppresses the oscillations that a nonzero $\delta m$ would cause, which is why it enters Eq.~\eqref{eq:dirac-condition}.

\subsection{Affleck-Dine realization}

The three conditions are met together in the leptoflavorgenesis of Ref.~\cite{Akita:2025zvq}, which turns the Affleck-Dine mechanism~\cite{Affleck:1984fy,Dine:1995kz} into a source of large flavor asymmetries: a scalar condensate stores the lepton charge in Q-balls through the sphaleron epoch and releases it at $T\gtrsim10\gev$, while medium effects still suppress the HNL interactions.  Beyond BBN, the dynamics may generate the baryon asymmetry of the Universe.  The large asymmetries can also produce sterile neutrino dark matter resonantly, in practice another HNL, lighter, at masses and mixings that a symmetric plasma cannot reach~\cite{Akita:2025txo}.

\emph{Baryon asymmetry.}  Affleck-Dine leptogenesis generates the baryon asymmetry together with the large late lepton asymmetries: most of the lepton charge is trapped in Q-balls that decay after the sphalerons have decoupled, while the small part released earlier is processed into baryons~\cite{Kawasaki:2002hq,Akita:2025zvq}.  For compensated flavor asymmetries, the conversion is further suppressed by the Yukawa couplings of the charged leptons~\cite{March-Russell:1999hpw,Mukaida:2021sgv}.  The flavor pattern of the early release and the decay temperature of the Q-balls select the sign and the size of the baryon asymmetry~\cite{Akita:2025zvq}.

\emph{Dark matter.}  A keV right-handed neutrino mixing with the active flavors is produced resonantly by large lepton asymmetries~\cite{Shi:1998km,Akita:2025txo}; at $L_\alpha\simeq0.1$, the production reaches mixing angles well below those of the symmetric case.  For the asymmetries that BBN and the CMB select, the resonance occurs at $T\simeq20$ to $80\mev$, after the HNLs freeze out and before they decay.  

The heavy HNLs add a new ingredient: the energy they release raises the comoving entropy by the factor $D=S_{\rm after}/S_{\rm before}$ and dilutes the light state, as heavier unstable species do in the $\nu$MSM~\cite{Asaka:2006ek}.  The transport bounds it: $D\simeq1$ to $7$ across the domains of Fig.~\ref{fig:money}, well below the enhancement by the asymmetries, so the small mixing angles survive.  The entropy release also lowers the momenta of the dark matter relative to the photon temperature by the factor $D^{1/3}$, that is, by $10$ to $50\%$, which makes the spectrum colder and weakens the bounds from structure formation on resonantly produced dark matter~\cite{Asaka:2006ek,Patwardhan:2015kga}.  At fixed asymmetry, resonant production grows linearly with the squared mixing angle~\cite{Akita:2025txo}, so its required value rises by $D$.  The dark matter number density stays below $1\%$ of the asymmetry, which therefore is not depleted.  A Dirac keV state is produced in the same way, and its radiative decay is half as fast as in the Majorana case~\cite{Abazajian:2017tcc}.  The dark matter band at asymmetries of this size lies beyond the range mapped in Ref.~\cite{Akita:2025txo} and follows from the same resonant-production calculation once the asymmetries are fixed.

\subsection{Model example}

The example addresses the complementary case, a generating mechanism that finishes while the sphalerons still operate.  The charge then waits in a singlet state, out of reach, and a heavy Dirac messenger $X$ delivers it to the active flavors once the conversion into baryons has shut off, decaying promptly into Standard Model leptons.  The HNL abundance and asymmetry are set only afterwards, by mixing, from the resulting plasma.  Because the $X$ decays create no new charge, the two quantities computed below, the yield it must carry and the entropy it adds, characterize the messenger alone.

\emph{Fields and transfer.}  Impose a global lepton number $U(1)_L$ that is exactly conserved at the effective-theory level, with
\begin{center}
\begin{tabular}{c|ccccc}
 field & $L_\alpha$ & $H$ & $N$ & $X$ & $N_s$ \\
 \hline
 $U(1)_L$ charge & $+1$ & $0$ & $+1$ & $+1$ & $+1$
\end{tabular}
\end{center}
where $N$, $X$, and the dark matter state $N_s$ are singlet Dirac fermions with identical quantum numbers.  The relevant terms are
\begin{align}
 \mathcal L\supset{}&
 -y_\alpha\overline L_\alpha\widetilde H P_RN -y'_\alpha\overline L_\alpha\widetilde H P_RX -y^s_\alpha\overline L_\alpha\widetilde H P_RN_s -m_N\overline NN-m_X\overline XX{}-m_s\overline N_sN_s.
 \label{eq:late-transfer-lagrangian}
\end{align}
Here, $P_R=(1+\gamma_5)/2$. The three singlets have identical quantum numbers, so the general mass matrix contains off-diagonal Dirac terms.  We work in its mass basis, with $N$, $X$, and $N_s$ the Dirac eigenstates and $y_\alpha$, $y'_\alpha$, $y^s_\alpha$ their Yukawa vectors in that basis, $\alpha=e,\mu,\tau$.  In that basis the mass matrix is diagonal, so no singlet couples to another directly.  Transitions such as $X\to NZ$ arise only through the mixing with the active neutrinos and are negligible at $|U|^2\lesssim10^{-8}$.

The three Yukawa vectors set different quantities.  The light Yukawa fixes the mixing that produces the HNL population, $U_\alpha\simeq y_\alpha v/(\sqrt2 m_N)$, with $v\simeq246\gev$ the Higgs vacuum expectation value.  The exact $U(1)_L$ forbids Majorana masses, so this mixing leaves $N$ Dirac and its lepton charge conserved.  The heavy Yukawa fixes how fast the messenger releases its asymmetry into the plasma.  At $m_X=100\gev$, the open channels are $X\to\ell W$ and $X\to\nu Z$, while $\nu h$ is closed, and the phase-space factors reduce the width to $\Gamma_X\simeq0.16\,y'^2m_X/(16\pi)$ with $y'^2=\sum_\alpha|y'_\alpha|^2$. 

The third Yukawa is the smallest: it puts the mixing of the dark matter state where resonant production needs it~\cite{Akita:2025txo}, and the symmetry keeps $N_s$ Dirac.  The nonresonant channel then populates it only negligibly, so no primordial component builds up, and the relic forms at the resonance alone, from the asymmetries.

The transfer, the release of the singlet charge into the plasma, faces two hazards.  The stored asymmetry exceeds the observed baryon asymmetry by nine orders of magnitude: of what reaches the active leptons while the sphalerons operate, no more than a fraction $10^{-9}$ is allowed.  And a nearly fully asymmetric population with yield of order $0.1$ carries energy comparable to the radiation, so its decays raise the entropy and dilute what they deliver.

Both hazards are avoided by releasing the charge late and in the right amount: it reaches the plasma only below the sphaleron freeze-out, either through an asymmetric $X$ population, made by a heavier singlet, or directly; how the asymmetry is generated stays unspecified.  The $X$ decay, faster than the expansion, finishes before the HNLs equilibrate, and requiring $X$ not to dominate it caps what it leaves at order $0.1$, exactly what the helium and $\neff$ conditions select.

Indeed, at $T_{\rm inj}=50\gev$ and $m_X=100\gev$, $y'\gtrsim10^{-7}$ does so within one Hubble time, well before the in-medium suppression of the mixing lifts.  With the yield $Y_{\Delta X}\equiv(n_X-n_{\bar X})/s=0.145$ and a degenerate Fermi gas, the most conservative spectrum, the energy fraction is $\rho_X/\rho_R=\tfrac43\,Y_{\Delta X}\langle E_X\rangle/T=0.89$, the entropy grows by $\Delta_{S,X}=(1+\rho_X/\rho_R)^{3/4}=1.6$, and $0.09$ survives.  Prompt and complete thermalization raises the plasma to $T_{\max}=T_{\rm inj}(1+\rho_X/\rho_R)^{1/4}=58.6\gev$, and both temperatures stay below the sphaleron freeze-out,
\begin{equation}
 T_{\rm inj}<T_{\rm sph},\qquad T_{\max}<T_{\rm sph}.
 \label{eq:model-sphaleron-condition}
\end{equation}
No baryon number is created; the entropy release only dilutes the conserved yield, $Y_B^{\rm before}=\Delta_{S,\rm tot}Y_B^{\rm obs}$, the $X$ step alone requiring $1.61\,Y_B^{\rm obs}$ before injection, and the HNL history raises that requirement accordingly.

The construction leaves no other trace.  Before the transfer, the same Yukawa makes a charge-symmetric $X$ component by freeze-in, $Y_X^{\rm FI}\simeq\tfrac{135}{8\pi^3\,1.66\,g_*^{3/2}}\,\Gamma_XM_{\rm Pl}/m_X^2\approx2\times10^{-3}$, with $M_{\rm Pl}$ the Planck mass and $g_*$ the effective number of relativistic degrees of freedom; unstable, it adds only a dilution of $1.004$.  What is released at $T_{\rm inj}$ decays into Standard Model leptons, so the asymmetry reaches the active flavors in full, with no residue and no additional scalar; the kinetic calculation of SuM~\ref{app:production} then builds the HNL population from the resulting plasma.  Before electroweak symmetry breaking, the couplings $y_\alpha$ also make the opposite helicities, a charge-symmetric early component whose effect on $\yp$ and $\neff$ is undone by shifts of at most $0.02$ along the family.

$Y_{\Delta X}$ is an input; an exact $U(1)_L$ cannot create net charge from a neutral state.  Its flavor pattern follows $y'_\alpha$ at injection, $Y_{L_\alpha}=(Y_{\Delta X}/\Delta_{S,X})\,{\rm Br}_\alpha$ with ${\rm Br}_\alpha\simeq|y'_\alpha|^2/\sum_\beta|y'_\beta|^2$, while the later division among the surviving HNLs, active flavors, and charged leptons comes from the coupled chemical and kinetic history of SuM~\ref{app:production} and~\ref{app:framework}.  A single asymmetric messenger gives all three flavor asymmetries one common sign, its branching fractions being positive, so a large compensated direction requires a second copy of $X$ with reversed charge; the Affleck-Dine release needs none~\cite{Akita:2025zvq}.  The example thus establishes a baryon-safe transfer of the right size; which primordial quantity sets which $L_\alpha$ stays undetermined.

\subsection{Comparison with seesaw models}

In seesaw models, where the HNLs generate the active neutrino masses, the number of HNL flavors decides which mixing patterns the oscillation data allow, and with them how much of the range below the kaon mass the combined constraints exclude~\cite{Boyarsky:2020dzc,Bondarenko:2021cpc,Chrzaszcz:2019inj,Domcke:2020ety}.  In our realization, the active neutrino masses arise from further right-handed neutrinos with Dirac masses, so the oscillation data constrain those states and leave the HNL mixings free.  We take the HNL to couple to a single flavor, the case adopted in the sensitivity studies of accelerator searches~\cite{Beacham:2019nyx,deBlas:2025gyz}.  As long as $|U_e|^2$ or $|U_\mu|^2$ is nonzero, the lifetime bound that the asymmetries relax depends on the ratios of the mixings only through a logarithm~\cite{Boyarsky:2020dzc,Bondarenko:2021cpc}.

\emph{The $\nu$MSM.}  An example of a seesaw model extended by sterile neutrino dark matter is the $\nu$MSM: of its three right-handed neutrinos, two are Majorana and quasi-degenerate, through an approximate lepton number symmetry, and the oscillations of this quasi-Dirac GeV pair generate late lepton asymmetries~\cite{Asaka:2005pn,Canetti:2012kh,Ghiglieri:2020ulj}.  Within the $\nu$MSM, Ref.~\cite{Drewes:2026tpy} shows that the asymmetric plasma, a birefringent medium for that pair, can amplify these asymmetries to yields of order $10^{-3}$, enough for resonant production of keV dark matter at the small mixing angles unconstrained by the current X-ray observations, and it ties this relic to GeV HNLs.  Large lepton asymmetries together with GeV HNLs are also the ingredients of the present scenario.

The $\nu$MSM solutions of Refs.~\cite{Canetti:2012kh,Ghiglieri:2020ulj,Drewes:2026tpy} do not realize the mechanism studied here, for two independent reasons.  The solutions of Table~\ref{tab:benchmarks} need flavor asymmetries $|L_\alpha|\simeq0.1$ to $0.3$, about a hundred times the largest yields, those of Ref.~\cite{Drewes:2026tpy}.  The splitting between the two quasi-degenerate HNLs, around $10^{-10}\,$eV at $m_N=1\gev$ and no smaller than about $4\times10^{-11}\,$eV at $2\gev$~\cite{Drewes:2026tpy}, exceeds the bound of Eq.~\eqref{eq:dirac-condition} by about 1.5 to 3.5 orders of magnitude, so the pair oscillates between $N$ and $\bar N$ thousands of times during its lifetime and the charge of the injected pions averages out.  The self-amplification of Ref.~\cite{Drewes:2026tpy} is also absent here: it needs the $N\leftrightarrow\bar N$ transitions near $T\simeq0.2\gev$, which violate lepton number, and the exact symmetry of our realization forbids them.  The asymmetries entering the calculation are therefore the primordial ones.

\section{Supernova bounds}
\label{app:sn}

Core-collapse supernovae constrain HNLs through the energy they carry out of the proto-neutron-star core, through their decays outside the star, and through the energy their decays deposit inside the envelope~\cite{Carenza:2023old,Chauhan:2023sci}.  Ref.~\cite{Carenza:2023old} treats Dirac HNLs mixing with the muon flavor over $10\mev\lesssim m_N\lesssim600\mev$, the case considered here, on an $18\,M_\odot$ progenitor whose core peaks at $40\mev$.  Ref.~\cite{Chauhan:2023sci} treats the energy-deposition constraint for Majorana HNLs mixing with the tau flavor over $100$ to $500\mev$, and argues that a similar constraint holds for muon mixing, where charged-current production off muons adds further channels.

The contour of Ref.~\cite{Carenza:2023old} drawn in Fig.~\ref{fig:money} covers the domains of Fig.~\ref{fig:money} at $m_N\lesssim0.4\gev$, which includes the region DUNE and Hyper-K would reach.  At larger masses the exclusion ends quickly, and where it ends is the part of the calculation most sensitive to the supernova model.  For $m_N$ well above the core temperature, production sits in the Boltzmann tail, $\Gamma_N\propto e^{-m_N/T_{\rm eff}}$, on top of the dependence on density, chemical potentials, radius, and trapping that the full calculation carries.

Ref.~\cite{Chauhan:2023sci} shows the progenitor dependence directly.  Its $18.8\,M_\odot$ model reaches core temperatures of about $40\mev$ and its $8.8\,M_\odot$ model about $30\mev$, and the two give substantially different constraints.  The luminosity fits given there carry the Boltzmann factor $e^{-m_N/T}$, normalized at those two temperatures, alongside powers of the temperature and of the nucleon degeneracy.  At $m_N=0.2\gev$, that factor alone separates the two models by about five, compared with about seven in the full calculation.  The same reference finds gravitational trapping to matter only for the heavier HNLs, whose spectrum peaks near $m_N$ and which therefore lack the kinetic energy to leave the potential well, and it reports the constraint weakening at those masses for that reason.

Where the constraint meets the SHiP sensitivity, it is most uncertain.  The core temperature varies between $30$ and $40\mev$ across the two progenitors of Ref.~\cite{Chauhan:2023sci}, and at $m_N=0.6\gev$, that alone changes the rate at which HNLs are produced by two orders of magnitude.  Ref.~\cite{Carenza:2023old} takes the upper value, so the contour of Fig.~\ref{fig:money} is the strongest of that range.  The heaviest mass the bound reaches is set by a fixed amount of Boltzmann suppression and scales with the temperature, so its position near $0.6\gev$ can move by $O(0.1\gev)$.  This is not a recalculation of the constraint, which also carries the radial profiles, the equation of state, the trapping, and the decay geometry, and which Ref.~\cite{Chauhan:2023sci} computes for Majorana HNLs mixing with the tau flavor.  Much of the overlap between the contour and the region SHiP can reach changes when that end moves.  A dedicated reanalysis of the gamma-ray channels finds the SN~1987A and diffuse constraints weaker than earlier estimates~\cite{Chauhan:2025mnn}.

Where laboratory searches and supernova constraints overlap, Ref.~\cite{Carenza:2023old} finds the energy deposited by the decays to dominate.  Its reach therefore carries the progenitor temperature and density profiles, the adopted core and envelope radii, the trapping and decay geometry, and the map from deposited energy to the observed explosion energy.  The two progenitors of Ref.~\cite{Chauhan:2023sci} already give constraints that differ substantially.  Neither analysis follows HNL production and decay inside a core-collapse simulation that responds to the HNL population.

We draw that contour hatched in Fig.~\ref{fig:money}, to mark that its high-mass end depends on the supernova model.  Even at that position, a substantial connected part of our domain within the reach of SHiP lies outside it.  A dedicated reassessment of these astrophysical bounds is left for future work.

\section{The QCD transition epoch and gravitational waves}
\label{app:qcd}

At the QCD transition epoch, $T\sim100$ to $200\mev$, the primordial flavor asymmetries have not yet been redistributed by the oscillations, which set in below $T\simeq15\mev$.  Large opposite-sign flavor components can therefore coexist there with a small total lepton charge, as in Affleck-Dine leptoflavorgenesis~\cite{Akita:2025zvq}.  

Electric neutrality requires the quarks to balance the net charge of the electrons, muons, and tau leptons.  The electron asymmetry thus acts on the quarks, while the tau asymmetry sits almost entirely in $\nu_\tau$ and reaches them only through the tau leptons, which are Boltzmann suppressed.  In the free-quark gas that enters the first-order criterion below, $L_\tau=0.1$ moves the larger of the magnitudes of the up and down quark chemical potentials, $\max(|\mu_u|,|\mu_d|)$, by $1.7\times10^{-4}$ of the change that $L_e=0.1$ produces at $T=125\mev$.  This fraction rises with temperature, to $1.7\times10^{-3}$ at $150\mev$ and $3\%$ at $200\mev$.  The hadron-resonance gas, in which we evaluate pion condensation below the crossover, therefore omits the tau leptons.  Shifting part of the net asymmetry between the electron and tau flavors changes the quark and electric-charge chemical potentials at the QCD transition epoch while leaving $\yp$, D/H, and $\neff$ almost unchanged.  Two realizations that BBN and the CMB cannot tell apart can thus behave differently at the QCD transition.  The charge-redistribution stage of SuM~\ref{app:production} computes $\mu_u(T)$, $\mu_d(T)$, and $\mu_Q(T)$ along the cooling trajectory at fixed comoving asymmetries.

Large flavor asymmetries can change the order of that transition.  For the compensated pattern $L_e=0$, $L_\mu=-L_\tau$, a two-flavor mean-field quark-meson analysis finds that $|L_\mu|\gtrsim0.1$ drives the cooling plasma into a pion-condensed phase through a first-order transition, with the latent heat growing rapidly with the asymmetry~\cite{Ferreira:2025zeu}.  At nonzero charge chemical potential, the critical endpoint of Ref.~\cite{Gao:2021nwz} also comes within reach.  A first-order transition at this epoch would source a stochastic gravitational wave background in the band of $\mu$Ares-class detectors, which makes the QCD transition epoch a target here.

Two routes could produce the signal: entry into the pion-condensed phase, and the chiral route, in which the quark chemical potentials pass above the critical endpoint.  The first is realized: along the family of asymmetries studied here, the plasma does enter it, first order in the mean-field analysis of Ref.~\cite{Ferreira:2025zeu}, and this defines the GW target region inside the $\neff$ deficit region.  The second is not: with the endpoint of current functional QCD calculations, no scanned point passes it, and where the confinement transition has been analyzed, the walls are slow and the signal far below $\mu$Ares-class sensitivity~\cite{Cline:2025bwe}.

\subsection{Pion condensation}

Lattice QCD at nonzero isospin density places pion condensation at $T\lesssim160\mev$ once the magnitude of the charge chemical potential, $|\mu_Q|$, exceeds the pion mass $m_\pi$~\cite{Vovchenko:2020crk}.  For each solution of the scan, we compute $\mu_Q(T)$ of the confined plasma in the hadron-resonance gas, in which $|\mu_Q|$ is largest just below the crossover and falls as the temperature drops.  We define the condensation margin as $\max|\mu_Q|/m_\pi$ over $T\leq140\mev$, well below the crossover, and take a value of one or more to mark entry into the condensed phase.  For compensated flavor asymmetries, Ref.~\cite{Vovchenko:2020crk} finds the onset at $|L_e+L_\mu|\gtrsim0.1$.  Our hadron-resonance gas reproduces this threshold at $T\simeq158\mev$ and requires $|L_e+L_\mu|\simeq0.13$ at $140\mev$.  In the mean-field quark-meson analysis of Ref.~\cite{Ferreira:2025zeu}, the plasma enters the condensed phase through a first-order transition.  The condensation criterion involves neither the critical endpoint nor the calibration of the quark chemical potentials introduced below.  The criterion is approximate: the hadron-resonance gas is ideal, and $\mu_Q/T$ approaches unity at these temperatures.  The hadron-resonance gas and a lattice-susceptibility description give charge chemical potentials that agree to about ten percent (SuM~\ref{app:validation}).  \par Together with the $\neff$ deficit window, the condensation criterion defines the GW target region of Fig.~\ref{fig:money}.  The region spans all nine masses, and a substantial part lies within the SHiP reach.  Evaluating the condensation margin up to the crossover, as in Ref.~\cite{Vovchenko:2020crk}, enlarges the region by about a tenth.

Whether the entry produces an observable background depends on the bubble-wall dynamics, which nothing published covers here.  The wall-velocity and signal conclusions of Ref.~\cite{Cline:2025bwe} apply to the weak, small-supercooling branch that it studied, and the one calculation that exists for the condensed phase finds the latent heat growing rapidly with the asymmetry~\cite{Ferreira:2025zeu}, so what evidence there is points to a stronger transition exactly where the asymmetries are largest.  A recent local-thermal-equilibrium scalar-fluid analysis also indicates that the obstruction to fast walls can be weaker than its steady-state estimate, allowing detonation or runaway solutions~\cite{Laurent:2026jvx}, although it does not treat QCD.  A first-order transition is not even necessary: at large tau flavor asymmetries the speed of sound in the condensed phase may exceed its conformal value, leaving an imprint on the low-frequency spectrum from causal sources~\cite{Ferreira:2026uzi}.

A complete prediction requires the equation of state at these chemical potentials, a nucleation calculation determining the transition temperature and the inverse duration of the transition in Hubble units $\beta/H$, a check that the transition completes, and the wall dynamics.  We therefore provide a concrete gravitational wave target without an amplitude, which we leave for future work.

\subsection{The chiral transition}

To compare our ideal-gas quark potentials with the normalization in which the endpoint of Ref.~\cite{Gao:2021nwz} is quoted, define
\begin{equation}
 \mu_*(T)\equiv3.05\,\max\left(|\mu_u(T)|,|\mu_d(T)|\right),
 \label{eq:mu-star}
\end{equation}
where the factor is fixed once, by calibrating the ideal-gas potentials against the Dyson-Schwinger endpoint of that reference.  Ref.~\cite{Gao:2021nwz} tabulates the flavor asymmetry at which the transition turns first order along several directions in the space of flavor asymmetries.  We evaluate our ideal-gas potentials on those same directions, at their quoted thresholds, and take the factor that carries our $\max(|\mu_u|,|\mu_d|)$ onto their endpoint value $\mu_{\rm CEP}$.  Both sides of that ratio are chemical potentials, which Eq.~\eqref{eq:fo-criterion} compares.  The four directions along which their calculation and ours both cross give $2.97$, $3.05$, $3.05$, and $3.14$, with our potentials evaluated at the entropy density of the cooling plasma in our solutions, and we adopt the mean.  The transition is a crossover at chemical potentials below the endpoint and first order above it.  With the endpoint at $(T_{\rm CEP},\mu_{\rm CEP})=(125,111)\mev$, obtained in the rainbow-ladder truncation of the Dyson-Schwinger equations, where $\mu_{\rm CEP}$ denotes the chemical potential of the light quarks, a trajectory is classified as reaching the first-order regime when
\begin{equation}
 \mu_*(T_{\rm CEP})\geq1.2\,\mu_{\rm CEP}.
 \label{eq:fo-criterion}
\end{equation}
The $20\%$ margin is an overshoot we adopt because the calibration above scatters by about six percent across the four directions, fixed by no QCD result; the continuous value of $\mu_*(T_{\rm CEP})/\mu_{\rm CEP}$ is recorded for every point.  The criterion establishes that the computed trajectory reaches that regime, at the endpoint's own temperature; the exit temperature below locates where it leaves the symmetric phase.  Estimating the strength of the transition requires nucleation and wall-dynamics calculations that we do not perform.

Current functional QCD calculations place the endpoint at about $(118,200)\mev$ in the same normalization, which corresponds to a baryon chemical potential near $600\mev$~\cite{Fu:2019hdw,Gao:2020fbl,Gunkel:2021oya,Gao:2024fhm}.  With this endpoint, no solution within the BBN and CMB windows reaches the first-order regime.  This agrees with Ref.~\cite{Gao:2024fhm}, where a first-order transition requires flavor asymmetries about three times larger than in Ref.~\cite{Gao:2021nwz}.  With the endpoint of Ref.~\cite{Gao:2021nwz}, a part of the $\neff$ deficit region meets Eq.~\eqref{eq:fo-criterion} and has, for at least one slope of Eq.~\eqref{eq:fo-line-family}, an exit temperature below the range analyzed in Ref.~\cite{Cline:2025bwe}.  Reaching the first-order regime therefore depends on where the endpoint lies, and the GW target region is defined by the entry into the pion-condensed phase.

For the first-order criterion, the location of the critical surface at nonzero charge chemical potential is the principal external QCD input, and the calibration factor, the $20\%$ margin, the slope family, and the corridor temperature range below are all ours.  The trajectory itself carries an equation-of-state uncertainty: a lattice-susceptibility description of the same region gives a charge chemical potential about ten percent larger than the free-quark and hadron-resonance-gas description used here (SuM~\ref{app:validation}), which would raise the margin of Eq.~\eqref{eq:fo-criterion}.

\emph{Exit temperature and the slow-wall range.}  The slope of the first-order line away from the endpoint is not known, so we take a family of lines anchored there,
\begin{equation}
 \mu_c(T;k)=\mu_{\rm CEP}\left[1+k\left(1-\frac{T}{T_{\rm CEP}}\right)\right], \qquad k=0,\,2,\,5,
 \label{eq:fo-line-family}
\end{equation}
with $k$ an assumed model input.  For each slope, the exit temperature $T_c(k)$ is the temperature at which the trajectory crosses the corresponding line, $\mu_*(T_c(k))=\mu_c(T_c(k);k)$, and it is where the realization leaves the symmetric phase.

Ref.~\cite{Cline:2025bwe} analyzes exit temperatures $84$ to $118\mev$, which we call the slow-wall corridor: an exit inside it falls in the range that reference covers, where the walls are slow and the signal suppressed.  A maximum supercooling of a few percent, predicted independently for SU($N$) Yang-Mills~\cite{Agrawal:2025xul}, points the same way, although that result is for the pure gauge theory at zero chemical potential, while the trajectories here carry dynamical fermions and $\mu_Q\simeq100$ to $300\mev$.  Each realization is classed in one of three ways.  Every slope may exit inside the corridor; at least one slope may exit below it; or the trajectory may enter the pion-condensed phase.  The last two both lie outside the range the existing calculation covers, and the evidence behind them differs.  For an exit below the corridor no wall-velocity calculation covers that range, and nothing is known about the signal in either direction.  The evidence for the pion-condensed entry is in the preceding subsection.  Because the slope is unknown, the escape condition is deliberately permissive: one of the three slopes exiting below $84\mev$ is enough, so such a point is a possible escape, which need not hold for every slope.  For every selected point we record $T_c$ for each slope, the condensation margin, and the margin on the first-order criterion.

\subsection{Primordial black holes}

The scenario predicts no primordial black holes: forming them at the QCD transition epoch needs a curvature power spectrum of order $10^{-3}$ at horizon scales~\cite{Gonin:2026tyu}, far above the amplitude the CMB measures, and nothing here sources such perturbations.  The connection runs the other way: lepton flavor asymmetries change the equation of state there, and with it the collapse threshold~\cite{Bodeker:2020stj,Gonin:2026tyu}.  These works consider compensated values of about $0.1$ and find a reduced softening, so fewer black holes form near a solar mass.  The asymmetries of our solutions are larger, and in the GW target region the plasma enters the pion-condensed phase, which neither study covers.

\section{Some results and discussion}
\label{app:detailed}

This appendix collects the detailed versions of the main-text results.  Table~\ref{tab:benchmarks-extended} lists the individual residuals of three reference solutions of the domains, complementing the benchmarks of Table~\ref{tab:benchmarks}.

\begin{table}[t!]
\caption{Individual residuals of three reference solutions of the domains; here, $\delta_D\equiv100\,\Delta(\mathrm{D/H})/(\mathrm{D/H})$, $R_\nu$ is the relic neutrino number ratio of Eq.~\eqref{eq:cnub-number-ratio}, and parentheses give the $1\sigma$ statistical uncertainty on the last digits.  The statistical uncertainties are those of the twelve-seed stress set.}
\label{tab:benchmarks-extended}
\begin{ruledtabular}
\begin{tabular}{cccccccc}
\# & $(m_N,\tau_N)$ & $q_N$ & $\xi_{20}$ & $\Delta\yp$ & $\delta_D[\%]$ & $\dneff$ & $R_\nu$ \\
\hline
1 & $(0.50\gev, 0.0484\s)$ & $+0.23$ & $(+1.72,-0.35,+1.77)$ & $-0.0017(9)$ & $-0.34(23)$ & $+0.003(2)$ & $1.455$ \\
2 & $(0.70\gev, 0.022\s)$ & $+0.72$ & $(-2.61,+0.62,+2.61)$ & $+0.0001(9)$ & $-0.60(34)$ & $-0.059(13)$ & $1.049$ \\
3 & $(0.60\gev, 0.143\s)$ & $+0.58$ & $(+2.71,-0.75,+3.61)$ & $-0.0012(8)$ & $-0.88(29)$ & $-0.061(10)$ & $1.645$ \\
\end{tabular}
\end{ruledtabular}
\end{table}

\begin{figure}[t!]
    \centering
    \includegraphics[width=0.7\linewidth]{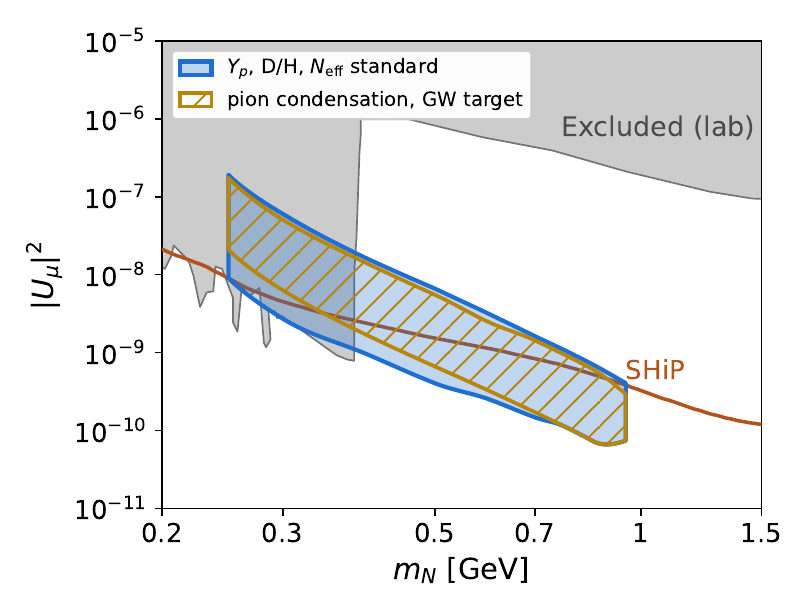}
    \caption{Relaxed selections on the axes of Fig.~1: the standard $\neff$ domain (blue) and the realizations for which some division of the asymmetry between the electron and tau flavors brings the QCD plasma into the pion-condensed phase (SuM~\ref{app:qcd}) while all BBN and CMB observables remain standard (gold).}
    \label{fig:gwinv}
\end{figure}

In the main text, the GW target domain includes the requirement of an $\neff$ deficit.  the GW target region sits inside the $\neff$ deficit region only to isolate realizations with a measurable deficit; the QCD condition itself does not involve $\neff$.  Here, we impose instead the no-imprint windows of the standard $\neff$ region: the same $\yp$ and D/H requirements as in Table~\ref{tab:domain-definitions}, together with $|\dneff|\leq0.05$.  Fig.~\ref{fig:gwinv} shows this relaxed selection on the axes of Fig.~\ref{fig:money}.  The gold band marks the grid points at which some division of the asymmetry between the electron and tau flavors keeps $\yp$, D/H, and $\neff$ inside the no-imprint windows and brings the QCD plasma into the pion-condensed phase.  As in the GW target region, the criterion for entering that phase is a condensation margin $\max|\mu_Q|/m_\pi$ of one or more at $T\leq140\mev$, where $\mu_Q$ is the charge chemical potential (SuM~\ref{app:qcd}).  This band extends over the whole mass range, and 21 of its 28 grid points lie within the sensitivity of SHiP.  An HNL discovered there could leave BBN and the CMB entirely standard, while the plasma still meets the QCD condition of the GW target domain.

\emph{The sign of $L_e$.}  This freedom includes the sign of the electron asymmetry, and both signs give viable solutions.  Those with $L_e<0$ keep the $\neff$ deficit window and still bring the QCD plasma into the pion-condensed phase, while the enhancement of the relic neutrino number density is lost: at $(0.7\gev,0.022\s)$, inside the SHiP reach, the asymmetries $(-0.202,+0.068,+0.142)$ give $\dneff=-0.11$ and a condensation margin of at least one, and the relic neutrino number density stands at $0.96$ of its standard value.  This solution has $L_e<0$.  Most of the GW target region within the SHiP reach lies on the branch with $L_e<0$.

\begin{figure}[t!]
    \centering
    \includegraphics[width=\linewidth]{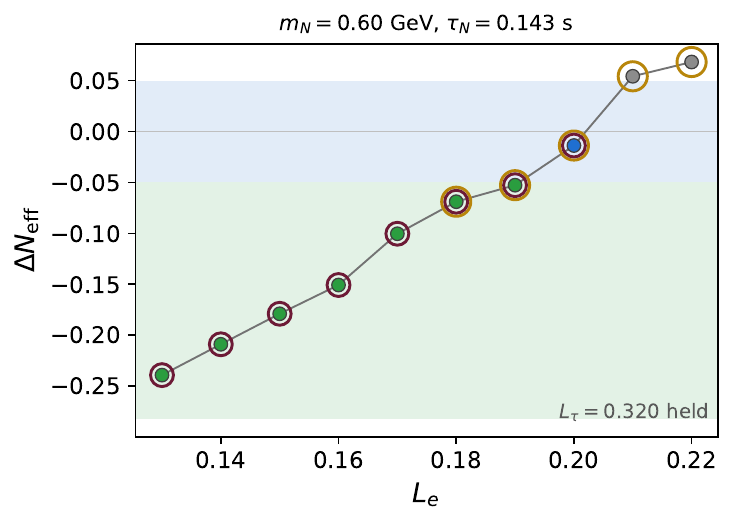}
    \caption{A trajectory in the division of the asymmetry between the electron and tau flavors, at $(m_N,\tau_N)=(0.60\gev,0.143\s)$.  The electron asymmetry increases along the curve from $L_e=+0.13$ to $+0.22$, the tau asymmetry is fixed at $L_\tau=0.320$, and $L_\mu$ is solved at every point from the helium condition, which keeps $\Delta\yp$ inside its window along the whole curve.  $\dneff$ is then an output, and it runs from $-0.24$ to $+0.07$ while D/H stays inside its window.  Point color gives the standard $\neff$ region, the $\neff$ deficit region, or neither.  Gold rings mark the points for which the QCD plasma enters the pion-condensed phase (SuM~\ref{app:qcd}); dark-red rings mark the points whose relic neutrino number density is enhanced by more than $30\%$.  The curve passes through the $\neff$ deficit domain, the onset of pion condensation, and the standard $\neff$ domain in turn, so which signature accompanies a discovery at this one mass and lifetime is decided by the position along the family.}
    \label{fig:sharing}
\end{figure}

\emph{The family of asymmetries at a single laboratory point.}  Figure~\ref{fig:sharing} walks it explicitly at $(m_N,\tau_N)=(0.60\gev,0.143\s)$, so the panel shows where that point sits inside its own family.  We hold $L_\tau$ at $0.320$, advance $L_e$ from $+0.13$ to $+0.22$ in steps of $0.01$, and re-solve $L_\mu$ from the helium condition at every step, so that $\dneff$ is an output.  It rises monotonically from $-0.239$ to $+0.068$, while $\delta_D$ stays inside its window throughout, between $-2.73$ and $+0.62$ percentage points.  The seven most electron-poor members lie in the $\neff$ deficit window.  For the two most electron-rich of them, at $L_e=0.18$ and $0.19$, the QCD plasma also enters the pion-condensed phase, as it does from $L_e=0.20$ to $0.22$.  At $L_e=0.20$, the family crosses into the standard $\neff$ window.  The relic neutrino enhancement grows monotonically along the whole walk, from $R_\nu=1.50$ to $1.74$, so every member inside a window also carries it.  At $L_e=0.19$, for example, the $\neff$ deficit, pion condensation, and $R_\nu>1.5$ coexist.  The asymmetries are free parameters of the scenario, so which signature accompanies a discovery at this mass and lifetime is set by where in the family the Universe sits.

\section{Arbitrary mixing patterns}
\label{app:mixing}

We solve explicitly only the case of mixing with a single flavor.  For a general pattern, the HNL production receives contributions from all three active flavors, weighted by the corresponding mixings together with the flavor-dependent collision rates and in-medium potentials, so the attractor of Eq.~\eqref{eq:production-boltzmann} is a combination of the three Fermi-Dirac distributions.  The weights are not the vacuum $|U_\alpha|^2$ alone: at the asymmetries reached here the matter potentials differ between flavors, and they enter the production rates alongside the mixings.  Only to first order in the degeneracies does such a combination have Fermi-Dirac form, with a chemical potential given by the weighted mean of the three.  At the values reached here, $|\xi_\alpha|\simeq1$ to $3$, no single effective degeneracy describes it, and we do not attempt a quantitative treatment of the general case.

A counting argument does carry over.  The helium and $\neff$ requirements remain two conditions on the three primordial flavor asymmetries.  Wherever a solution exists, and the two conditions are independent, they leave a one-parameter family of asymmetries, as in the single-flavor case of the main text.  Changing the mixing pattern deforms both conditions: through the flavor-dependent production rates, through the charge redistribution that maps $L_\alpha$ onto $\xi_\alpha$, and through the decay channels and the electromagnetic heating they produce.  It can therefore move the family, shrink it, or remove it altogether where the two conditions cease to have a common solution.

The two conditions are not carried by the same flavors in every pattern.  For muon mixing, $\xi_\mu$ fixes the particle-antiparticle ratio the HNLs acquire at production, while $\xi_e$ shifts the weak $\pn$ equilibrium the mesons act on, so the two effects come from different flavors.  For electron mixing, one degeneracy does both.  Naively, the electron case would then be insecure, with a single degeneracy asked to produce both effects at once.

The oscillations separate the two effects in time.  The particle-antiparticle ratio is fixed when the HNLs decouple from the plasma, at $T\simeq0.25$ to $2\gev$ (SuM~\ref{app:production}).  At $20\mev$, where the production stage ends, the flavor charges are still strongly unequal: at the benchmark of Fig.~\ref{fig:asym-evolution} they stand in the ratio $0.37:-0.02:0.65$ for $e:\mu:\tau$ as fractions of the total.  By $10\mev$ the oscillations have driven them to $0.37:0.31:0.32$, so the composition present at production does not reach the epoch where the weak $\pn$ rates act, and the primordial $L_e$ is not the charge that shifts those rates for any pattern.

The charges do not stay equipartitioned.  At $1\mev$ they read $0.25:0.40:0.36$.  Several effects act between the two temperatures, and one of them is the decays: they return the asymmetry $q_N$ the HNL population carried, and they do so below $5\mev$, after the oscillation stage, so the oscillations do not undo what they deposit.  Where it lands follows from the chains.  For $N\to\pi^+\mu^-$ the electron-flavor injections of the two muon decays cancel and the net deposit sits in the muon channel, while for $N\to\pi^+e^-$ it sits in the electron channel.  Under electron mixing the charge that shifts the weak rates therefore carries a component set by $q_N$, and the two effects are linked again, later than at production and by a different route.

The coupling that remains is weak, and the scan measures it.  Changing the division of the asymmetry between the electron and tau flavors with $L_\mu$ re-solved so that helium is held fixed, which takes a muon adjustment of median $dL_\mu/dL_e=-0.38$, $\dneff$ still answers with median $d(\dneff)/dL_e=+1.8$ across the 179 families that carry at least three members.  Degeneracy of the two conditions would mean that the helium-preserving direction leaves $\dneff$ untouched, and it does not: a displacement $\delta L_e\simeq0.18$ carries $\dneff$ across the whole $0.33$ window.  That separation follows from what the two conditions test, the meson-driven conversion on one side and the neutrino energy density on the other.  It is measured under muon mixing, where the net decay deposit does not reach $\xi_e$; under electron mixing it does, so this figure bounds the separation there.  The two conditions cannot align in any pattern: helium responds to $\xi_e$ and to $r$, while $\neff$ responds to $\sum_\alpha\xi_\alpha^2$, the degenerate Fermi-Dirac energy density carrying no term linear in $\xi$.  No redistribution among the flavors turns one of those into a function of the other.

The microscopic mechanism itself is not equally robust across patterns.  It needs a charge-correlated charged-current meson channel, $N\to\pi^+l_\alpha^-$ with the conjugate for $\bar N$.  That channel opens at $m_N\gtrsim m_{l_\alpha}+m_\pi$, which is $0.14\gev$ for the electron and $0.25\gev$ for the muon, both inside the range studied here, but $1.9\gev$ for the tau, above it.  The modes that remain open for a tau-mixed HNL below that threshold, such as $N\to\nu_\tau\pi^0$, are neutral-current and carry no correlation between meson charge and lepton number.  An electron or muon admixture therefore preserves the mechanism once its own channel is open, while a growing tau admixture dilutes it, and a tau-dominated pattern removes it altogether in this mass range.

A tau admixture dilutes the mechanism through the injected charge.  Below the threshold its hadronic decays are neutral-current: $N\to\nu_\tau\pi^0$ injects no charged pions at all, and modes such as $N\to\nu_\tau\pi^+\pi^-$ inject them in pairs.  Writing $f$ for the charge-correlated fraction of the injected charged pions, the ratio the nucleons see is
\begin{equation}
    \frac{n_{\pi^+}}{n_{\pi^-}}=\frac{f\,r+(1-f)(r+1)/2}{f+(1-f)(r+1)/2}\,,
    \label{eq:pion-dilution}
\end{equation}
which equals $r$ at $f=1$ and unity at $f=0$.  Reaching a given neutron fraction then needs a larger HNL asymmetry than Eq.~\eqref{eq:rreq} gives, and the requirement grows as $f$ falls.  The viable mass window is in that sense pattern dependent, although a tau-dominated pattern is not the generic expectation.

For the electron and muon patterns that do keep the mechanism, production needs a degeneracy in the mixing flavor large enough to give $r\simeq r_{\rm req}$.  Across the accepted solutions the muon degeneracy required lies in $-0.7\leq\xi_\mu\leq2.6$, while the electron degeneracies realized along the same families span $-3.9\leq\xi_e\leq3.9$ and cover that interval completely.  A pattern that sourced the HNL asymmetry from the electron flavor would therefore ask for a degeneracy the families already reach, at a different member of the same family.  The computed $r$ ranges from $1.4$ to above $200$, so the estimate $r_{\rm req}\simeq4$ to $10$ of Eq.~\eqref{eq:rreq} is bracketed from both sides.  We therefore expect a change of pattern between the electron and muon flavors to move the families and to leave them intact.

Whether a pattern admits a solution at all is not settled by counting the free asymmetries.  Counting fixes the dimension of a solution set once that set is known to be non-empty, and the map from the primordial $L_\alpha$ to the observables is deformed by the pattern at every stage: the production weights, the charge redistribution, the branching into charge-correlated channels, and the division of the released energy between the neutrino and the electromagnetic sectors.  We therefore solved the coupled system explicitly at one node, $(0.5\gev,0.0484\s)$, on a quarter grid of the mixing simplex.  Thirteen patterns solved, and every one of them reached both the standard $\neff$ and the $\neff$ deficit window, from the single-flavor cases to those with three quarters of the mixing in tau.  At this node neither region is selective in the mixing, and the pattern acts inside the family, leaving its existence untouched.

The relic neutrino signature varies with the pattern, and it is selected by the net charge the family carries, which no single flavor fixes on its own.  Scanning the mixing simplex on a quarter grid at this node, thirteen patterns solve, and across them the enhancement follows the total charge of the accepted solution: $R_\nu-1\simeq(2$ to $3)\,|\sum_\alpha L_\alpha|$, from $|\sum_\alpha L_\alpha|=0.14$ to $0.21$.  The patterns differ in how much net charge the helium condition leaves behind.

The variation is set by how many flavors are mixed.  With one or two active flavors the accepted solutions carry $|\sum_\alpha L_\alpha|=0.14$ to $0.21$ and reach $R_\nu=1.28$ to $1.48$, the single-flavor cases being the largest at $1.48$.  With all three active the helium condition admits a nearly traceless combination, and every three-flavor pattern solved does exactly that, with $|\sum_\alpha L_\alpha|\simeq0.02$ and a relic-sea enhancement within about $2\%$ of standard; the $1{:}1{:}1$ pattern, for example, settles at $L=(0.044,-0.243,0.220)$.  The relic neutrino number density thus remains essentially standard.  A traceless family leaves no net charge for the relic neutrinos to inherit, and the third flavor makes tracelessness attainable while helium is still satisfied.

This is distinct from the tau threshold above, and it is not a tau effect: patterns with a large tau fraction keep the signature as long as a third flavor is absent, $|U_e|^2:|U_\tau|^2=1:3$ reaching $R_\nu=1.385$ and $1:1$ reaching $1.428$.  The standard $\neff$ and $\neff$ deficit windows are untouched by any of this.  Every pattern that solved reached both of them, including those with three quarters of the mixing in tau.  These are single-node results and the shift a pattern induces is itself node dependent.

Helium can always be restored by re-solving the asymmetry of the mixing flavor, so the $\neff$ condition decides whether a pattern keeps a solution, and the scan bounds the shift of $\dneff$ that the family can absorb.  Along a family $\dneff$ is an output that sweeps continuously, and its span across the members of one family has median $0.73$, more than twice the width $0.33$ of the union of the standard $\neff$ and $\neff$ deficit windows.  The span exceeds that width at $72\%$ of the sampled points, and the family crosses the window at $70\%$ of them, so the crossing is generic.  Changing the pattern acts on this picture mainly by shifting the whole $\dneff$ curve, since it redistributes the decay energy between the neutrino and the electromagnetic sectors.  A crossing family survives a rigid shift of $0.22$ at the median point and of $0.58$ in the upper quartile.  Patterns $|U_e|^2:|U_\mu|^2:|U_\tau|^2$ that shift $\dneff$ by less than that are absorbed by moving along the family.  A tau-dominated pattern is not in that class, since it removes the charge-correlated channel outright.  Locating the families and sizing the domains for a general pattern still requires the full calculation, which we leave for future work.

\clearpage
%
\begingroup
\makeatletter\@fileswfalse\makeatother
\InputIfFileExists{\jobname.bbl}{}{}
\endgroup

\end{document}